\documentclass[apj]{emulateapj}
\usepackage{apjfonts}

\slugcomment{Accepted for publication in The Astrophysical Journal}
\shorttitle{TURBULENT MIXING IN ACCRETING NEUTRON STARS}
\shortauthors{PIRO \& BILDSTEN}

\newcommand{\be}{\begin{eqnarray}}
\newcommand{\ee}{\end{eqnarray}}
\newcommand{\lp}{\left(}
\newcommand{\rp}{\right)}
\newcommand{\wa}{\omega_{\rm A}}

\newcommand{\wk}{\Omega_{\rm K}}

\newcommand{\dens}{\rho_6}
\newcommand{\temp}{T_8}
\newcommand{\mass}{M_{1.4}}
\newcommand{\radius}{R_{6}}
\newcommand{\spin}{\Omega_{0.1}}
\newcommand{\mdot}{\dot{m}_{0.1}}
\newcommand{\opac}{\kappa_{0.04}}
\newcommand{\tmix}{t_{\rm mix}}
\newcommand{\tacc}{t_{\rm acc}}
\newcommand{\tnuc}{t_{3\alpha}}

\newcommand{\yacc}{y_{\rm acc}}
\newcommand{\ymix}{y_{\rm mix}}

\begin{document}


\title{Turbulent Mixing in the Surface Layers of Accreting Neutron Stars}

\author{Anthony L. Piro\altaffilmark{1} and Lars Bildsten} 

\affil{Kavli Institute for Theoretical Physics, Kohn Hall, University
of California, Santa Barbara, CA 93106;\\ piro@kitp.ucsb.edu, bildsten@kitp.ucsb.edu}

\altaffiltext{1}{Current address: Astronomy Department and Theoretical Astrophysics Center,
601 Campbell Hall, University of California,
Berkeley, CA 94720; tpiro@astro.berkeley.edu}


\begin{abstract}
   During accretion a neutron star (NS) is spun up as angular momentum is
transported through its surface layers. We study the resulting differentially
rotating profile, focusing on the impact this has for type I X-ray bursts. The
predominant viscosity is likely provided by the Tayler-Spruit dynamo,
where toroidal magnetic field growth and Tayler instabilities balance to
support a steady-state magnetic field. The radial and azimuthal components
have strengths of $\sim10^5\ {\rm G}$ and $\sim10^{10}\ {\rm G}$, respectively.
This field provides a Maxwell stress on the shearing surface layers, which
leads to nearly uniform rotation at the depths of interest for X-ray bursts (near
densities of $\approx10^6\ {\rm g\ cm^{-3}}$). A remaining small shear
transmits the accreted angular momentum inward to the NS interior. Though
this shear gives little viscous heating, it can trigger turbulent mixing. Detailed
simulations will be required to fully understand the consequences of mixing,
but our models illustrate some general features. Mixing has the greatest
impact when the buoyancy at the compositional discontinuity between
accreted matter and ashes is overcome. This occurs at high accretion rates,
at low spin frequencies (when the spin is small, the relative speed of the
accreted material is larger), or may depend on the ashes from the previous burst.
 We then find two new regimes of burning. The
first is ignition in a layer containing a mixture of heavier elements from the ashes.
If ignition occurs at the base of the mixed layer, recurrence times as short as
$\sim5-30\ {\rm minutes}$ are possible. This may explain the short recurrence
time of some bursts, but incomplete burning is still needed to explain these bursts'
energetics. When mixing is sufficiently strong, a second regime is found where
accreted helium mixes deep enough to burn stably, quenching X-ray bursts. We
speculate that the observed change in X-ray burst properties near one-tenth the
Eddington accretion rate is from this mechanism. The carbon-rich material produced
by stable helium burning would be important for triggering and fueling superbursts.
\end{abstract}

\keywords{accretion, accretion disks ---
	stars: magnetic fields ---
	stars: neutron ---
	X-rays: bursts ---
	X-rays: stars}


\section{Introduction}

   As neutron stars (NSs) in low mass X-ray binaries accrete material from
their companions they are expected to be spun up by this addition of
angular momentum, possibly becoming millisecond
pulsars \citep{bv91}. This suspicion has received support
by the discovery of accretion driven millisecond pulsars \citep{wv98},
as well as the millisecond oscillations seen during type I X-ray bursts
\citep{cha03}, the unstable ignition of the accumulating fuel
\citep[for reviews, see][]{lvt95,sb06,gal06}. The majority of the rotational
kinetic energy of the accreted material is dissipated at low densities
($\sim10^{-4}-10^{-1}\ {\rm g\ cm^{-3}}$) in the boundary layer
\citep[for example,][]{is99}. Nevertheless, angular momentum
must be transported into the NS interior if it is to be spun up,
even if at times the angular momentum could be radiated
as gravitational waves \citep{bil98b}. This transport implies a non-zero,
albeit small, shear throughout the outer liquid parts of the NS. Such shearing
may lead to viscous heating as well as chemical mixing at depths far below
the low density boundary layer where the
majority of the shearing occurs.

   Such a picture was previously investigated by \citet{fuj88,fuj93} for both
accreting NSs and white dwarfs. His studies demonstrated that the shearing
profile of a NS could be large enough that mixing through baroclinic instabilities
may be important at the depths of type I X-ray bursts.
Such a result is compelling because it may help to explain some of the remaining
discrepancies between the theory and observations of these bursts.
Though a simple limit cycle picture has been successful in qualitatively explaining the bursts'
primary characteristics, including their energies ($10^{39}-10^{40}\ {\rm ergs}$),
recurrence times (hours to days), and durations ($\sim 10-100\ {\rm s}$)
\citep{fhm81,bil98a}, outstanding problems still remain \citep{fuj87b,vpl88,bil00}.
Chief among these is the critical accretion rate required for stable accumulation
(which we discuss in more detail in \S \ref{sec:theend}). Another problem is
the occurrence of successive X-ray bursts with recurrence times as short as $\sim10$
minutes \citep[][and references therein]{gal06}, for which there has been speculation
that this could be due to mixing during accretion \citep{fuj87b}.

   In this present work we reassess the importance of angular momentum
transport and the resulting mixing. We find that the hydrodynamic instabilities
studied by \citet{fuj93} are insufficient to prevent strong shearing of magnetic fields.
This leads to generation of the Tayler-Spruit dynamo \citep{spr02}, where toroidal
field growth is balanced by Tayler instabilities to create a steady-state magnetic field,
which provides a torque on the shearing layers that is larger than any purely
hydrodynamic mechanism. The result is nearly uniform rotation and little viscous heating,
too little to affect either X-ray bursts or superbursts \citep[thermonuclear
ignition of ashes from previous X-ray bursts,][]{cb01,sb02}. Turbulent mixing is found
to be non-negligible and in some cases may mix fresh material with the ashes of
previous bursts. The key is whether the strong buoyancy due to the larger density of
the ashes below can be overcome. When mixing occurs, we find two new regimes of
burning for an accreting NS. The first is that fuel can be mixed down to depths necessary for
premature unstable ignition. The timescale for ignition of such bursts
is short enough ($\sim5-30\ {\rm minutes}$) to explain the short recurrence time bursts \citep{gal06}.
The second is that if material is mixed to sufficiently large
depths (and therefore temperatures) it can burn stably, ceasing X-ray bursts altogether.
Such a mechanism may be responsible for the quenching of X-ray
bursts seen at surprisingly low accretion rates for many atoll class NSs \citep{cor03}.
We derive an analytic formula that estimates what spins and accretion rates necessary
to transfer between these two regimes (eq. [\ref{eq:mdotcrit2}]).

\subsection{The Basics of Angular Momentum Transport}

\begin{figure}
\epsscale{1.2} 
\plotone{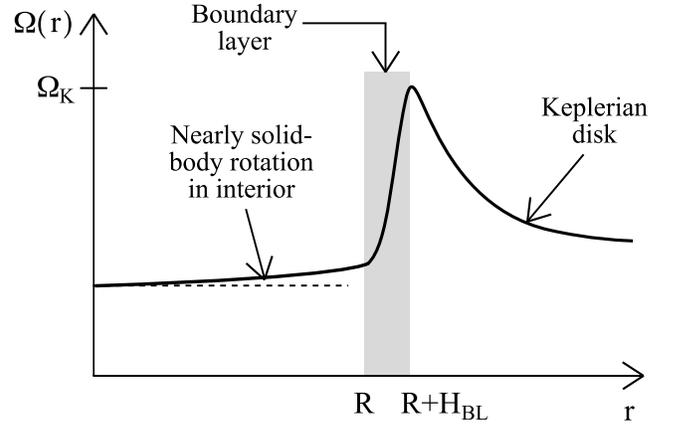}
\caption{A diagram demonstrating the rotation profile of material from the accretion disk,
through the boundary layer, and into the NS. Material reaches the NS surface with a Keplerian
spin frequency, $\wk$, the majority of which is dissipated within the boundary layer
({\it gray region}) with thickness $H_{\rm BL}\ll R$. Nevertheless, angular momentum is
still being added to the NS, and a torque must be communicated through the NS. This implies
a small, non-zero, amount of shear throughout the NS interior.}
\label{fig:rotation}
\epsscale{1.0}
\end{figure}

   Before beginning more detailed analysis, it is useful to present some general
equations for angular momentum transport. This demonstrates why we expect the NS
interior to be shearing and explicitly relates the torque 
from the accreting matter to this shear. Figure \ref{fig:rotation} schematically
shows the expected rotation profile from the accretion disk, through the boundary
layer, and into the NS interior. Material accreted at a rate $\dot{M}$ reaches the NS surface
with a nearly Keplerian spin frequency of
$\wk=(GM/R^3)^{1/2}=1.4\times10^4\ {\rm s^{-1}}\ \mass^{1/2}\radius^{-3/2}$, where
$M_{1.4}\equiv M/1.4M_\odot$ and $R_6\equiv R/10^6\ {\rm cm}$, which has a kinetic energy
per nucleon of $\approx200\ {\rm MeV\ nuc^{-1}}$. The majority of this energy
is dissipated in a boundary layer of thickness $H_{\rm BL}\ll R$
\citep[as studied by][]{is99} and never reaches far into the NS surface.
Nevertheless, angular momentum is being being added at a rate of
$\dot{M}(GMR)^{1/2}$, so that a torque of this magnitude must be communicated
into the NS. This implies a non-zero shear rate in the interior liquid layers, down to the
solid crust. In the present work we are interested in the shear at the depths where X-ray bursts
ignite, near $\rho\approx10^6\ {\rm g\ cm^{-3}}$, which is well below the boundary layer.
Though we focus on a magnetic mechanism for angular momentum transport, we
expect the boundary layer and densities up to $\approx5\times10^3\ {\rm g\ cm^{-3}}$
to be dominated by hydrodynamic instabilities (which we discuss in more detail
in \S \ref{sec:magnetic}).

   The pressure scale height at the depth of X-ray bursts is $H\approx30\ {\rm cm}\ll R$,
which allows us to use a Cartesian coordinate system
with $z$ as the radial coordinate. This is far below the depths considered by
\citet{is99} when they investigated the uneven covering of the NS surface by
accreted fuel. Furthermore, all of the transport
mechanisms we consider work most efficiently in directions perpendicular to gravity
(because no work is performed), thus it is a good approximation to consider the surface
layers as concentric spheres, each with constant $\Omega$. Transfer of angular momentum
is reduced to a one-dimensional diffusion equation \citep{fuj93}
\be
	\frac{d}{dt}(R^2\Omega) = \frac{1}{R^2\rho}\frac{\partial}{\partial z}
		\lp\rho \nu R^4 \frac{\partial\Omega}{\partial z}\rp,
\ee
where $\Omega$ is the NS spin frequency and $\nu$ is the viscosity.
The total time derivative is given by
$d/dt=\partial/\partial t+V_{\rm adv}\partial/\partial z$, where $V_{\rm adv}$ is the
advecting velocity of the fluid in an Eulerian frame.
In steady-state, we take $\partial/\partial t=0$
and set $V_{\rm adv}=-\dot{M}/(4\pi R^2\rho)$ so that
\be
	-\frac{\dot{M}}{4\pi R^2}\frac{d}{dz}(R^2\Omega) = \frac{1}{R^2}\frac{d}{dz}
		\lp\rho \nu R^4 \frac{d\Omega}{dz}\rp.
\ee
Integrating from the surface where the spin is $\Omega_{\rm K}$ (due to the disk)
down to a depth $z$ where the local spin is $\Omega$, and assuming
$d\Omega/dz=0$ at the surface,
\be
	-\dot{M}R^2\Omega_{\rm K}+\dot{M}R^2\Omega
	= -4\pi \rho\nu R^4 d\Omega/dz.
\ee
Taking the limit $\Omega\ll\Omega_{\rm K}$ we find,
\be
	4\pi R^3\rho\nu q\Omega = \dot{M}R^2\wk,
	\label{eq:angularmomentum}
\ee
where $q\equiv d\log\Omega/d\log z$ is the shear rate.
This equation shows that $q>0$ when angular momentum is transported
inward. In general, we will find that $q$ is rather small ($\lesssim1$)
at the depths of interest. Nevertheless $q$ is large enough to
activate instabilities that help to transport angular momentum,
as well as mix material.

\subsection{Outline of Paper}

   We begin by comparing and contrasting some well-known hydrodynamic
instabilities in \S \ref{sec:hydrodynamic} and estimate the resulting shear rates.
In \S \ref{sec:magnetic}, we discuss
the consequences that this shearing has on a magnetic field, which motivates
implementation of the Tayler-Spruit dynamo. In \S \ref{sec:nomix}, we
calculate accumulating NS envelopes without the effects of viscous
angular momentum transport. This enables us to judge when such effects
must be incorporated. We calculate models including mixing in \S \ref{sec:mix}.
We conclude in \S \ref{sec:theend} with a summary of our results and
a discussion of type I X-ray burst and superburst observations.

\section{Hydrodynamic Viscosity Mechanisms}
\label{sec:hydrodynamic}

   In the following sections we discuss hydrodynamic instabilities. This is not an
exhaustive survey \citep[for further details, see][]{heg00}, but rather is meant to
summarize those instabilities that are
most crucial to our problem, so as to set the context for the magnetic transport
mechanism we study later.


\subsection{Kelvin-Helmholtz Instability}
\label{sec:kelvinhelmholtz}

   The Kelvin-Helmholtz instability (also referred to as the dynamical
shear instability) is governed by the Richardson number
\be
	Ri \equiv \frac{N^2}{q^2\Omega^2},
	\label{eq:richardson}
\ee
where $N$ is the Brunt-V\"{a}is\"{a}l\"{a} frequency, which
is composed of contributions from thermal and compositional buoyancy,
\be
	N^2 = N^2_T+N^2_\mu.
\ee
The thermal buoyancy is given by
\be
	N^2_T = \frac{g}{H}\frac{\chi_T}{\chi_\rho}
		\left[\nabla_{\rm ad}-\lp\frac{d\ln T}{d\ln P}\rp_*\right],
\ee
where $g=GM/R^2=1.87\times10^{14}\ {\rm cm\ s^{-2}}\mass\radius^{-2}$ is the surface
gravitational acceleration (ignoring the effects of general relativity),
$\chi_Q=\partial\ln P/\partial\ln Q$, with all other intensive variables
set constant, $\nabla_{\rm ad}=(\partial\ln T/\partial\ln P)_{\rm ad}$
is the adiabatic temperature gradient,
the asterisk refers to derivatives of the envelope's profile, and the pressure
scale height is,
\be
	H=P/\rho g=33.1\ {\rm cm}\ \mu_{1.33}^{-1}\temp,
\ee
where $T_8\equiv T/10^8\ {\rm K}$. For these
analytic estimates we assume an ideal gas equation of state and
use a pure helium composition with a mean molecular weight
of $\mu=1.33\mu_{1.33}$. We omit the scalings with mass and radius to simplify presentation.
The compositional buoyancy is \citep{bc98}
\be
	N^2_\mu = -\frac{g}{H\chi_\rho}\left[ \chi_{\mu_e}\lp\frac{d\ln\mu_e}{d\ln P}\rp_*
		+ \chi_{\mu_i}\lp\frac{d\ln\mu_i}{d\ln P}\rp_* \right],
\ee
where $\mu_e$ ($\mu_i$) is the mean molecular weight per electron (ion).
Linear analysis shows that Kelvin-Helmholtz instability occurs
when $Ri<1/4$, which develops into strong turbulence that readily
transports angular momentum. This result assumes that thermal diffusion can
be ignored for the unstable fluid perturbations, in other words, that the perturbations
are adiabatic.

   Fluid perturbations with a characteristic size $L$ and speed $V$ become non-adiabatic
when the timescale for thermal diffusion, $L^2/K$, where $K$ is the thermal diffusivity,
is less than the timescale of the perturbation, $L/V$. The ratio of these two timescales is the
P\'{e}clet number, $Pe\equiv VL/K$ \citep{tow58}. The restoring
force provided by thermal buoyancy is weakened when $Pe<1$,
which requires the substitution of $N_T^2\rightarrow PeN^2_T$ and
promotes instability for regions where $N_T>N_\mu$.
Thermal diffusion is most efficient at small lengthscales, which motivates setting
$LV/\nu_k=Re_c$ \citep{zah92}, where $\nu_k$ is the kinematic viscosity and $Re_c$
is the critical Reynolds number for turbulence, which is of order 1000. This gives the
P\'{e}clet number approximately related to the Prandtl number, $Pr$, by
$Pe\approx Re_cPr$. The turbulent
perturbations are thus non-adiabatic when \citep{zah92}
\be
	K>\nu_k Re_c.
	\label{eq:secularcondition}
\ee
In the non-degenerate surface layers the kinematic viscosity is dominated
by ions, and has a value of \citep{spi62}
\be
	\nu_k =1.4\times10^{-3}\ {\rm cm^2\ s^{-1}}\rho_6^{-1}T_8^{5/2},
	\label{eq:ionviscosity}
\ee
where $\rho_6\equiv\rho/10^6\ {\rm g\ cm^{-3}}$,
and we assume a Coulomb logarithm of $\ln\Lambda=20$. Setting
$K =16\sigma_{\rm SB}T^3/(3c_p\kappa\rho^2)$, where
$\sigma_{\rm SB}$ the Stefan-Boltzmann constant, $c_p$ the specific heat,
and $\kappa$ the opacity, the thermal diffusivity is
\be
	K = 48.8\ {\rm cm^2\ s^{-1}}
		\mu_{1.33}\opac^{-1}\dens^{-2}
		\temp^3,
	\label{eq:thermal}
\ee
where we approximate $c_p=5k_{\rm B}/2\mu m_p$ and
scale the opacity to $\opac\equiv\kappa/0.04\ {\rm cm^{2}\ g^{-1}}$
\citep[the opacity is largely given by electron scattering, but is decreased by
degeneracy effects, see][]{pac83,bil98a}.
Substituting equations (\ref{eq:ionviscosity}) and (\ref{eq:thermal}) into
equation (\ref{eq:secularcondition}), we find that the
perturbations are non-adiabatic at depths of
$\rho~\lesssim~\mbox{$3\times10^7$}\ {\rm g\ cm^{-3}}\ T_8^{1/2}$.
The new ``secular'' Richardson number associated with this limit is,
\be
	Ri_s\equiv \frac{\nu_kRe_c}{K}\frac{N_T^2}{q^2\Omega^2}.
	\label{eq:secular}
\ee
When $Ri_s<1/4$, the so-called ``secular shear instability'' arises.

   The competing effects of accretion increasing $q$ versus turbulence
developing when $Ri_s<1/4$ (and decreasing $q$) drive the surface
layers toward marginally satisfying $Ri_s=1/4$ (assuming at this moment that the
sole viscous mechanism is the Kelvin-Helmholtz instability).
This expectation is borne out in the white dwarf studies of \citet{yl04}.
Thus we can trivially estimate the $q$ due to this mechanism.
The thermal buoyancy is \citep{bil98a}
\be
	N_T = \lp\frac{3}{20}\frac{g}{H}\rp^{1/2}=9.2\times10^5\ {\rm s^{-1}}
		\mu_{1.33}^{1/2}\temp^{-1/2}.
\ee
We substitute $Ri_s=1/4$ into equation (\ref{eq:secular}), and assuming
$Re_c=1000$, solve for a shear rate of
\be
	q_{\rm KH}=223\ \opac^{1/2}\dens^{1/2}\temp^{-3/4}
		\Omega_{0.1}^{-1},
	\label{eq:qkh}
\ee
where $\Omega_{0.1}=\Omega/0.1\Omega_{\rm K}$. A shear rate this
large would promote prodigious viscous heating, as well as ample
mixing, but as we soon show, such a large shear is prohibited by other instabilities.

\subsection{Baroclinic Instability}
\label{sec:baroclinic}

   Another important hydrodynamic instability that has been studied extensively for accreting
degenerate stars is the baroclinic instability \citep{fuj93}, which we quickly
summarize here. The interested reader should consult \citet{fuj87,fuj88} for
further details \citep[also see the discussion in][]{cb00}.

   The baroclinic instability arises because surfaces of constant pressure
and density no longer coincide if hydrostatic balance is to be maintained
when differential rotation is present. In such a configuration, fluid perturbations
along nearly horizontal directions are unstable, though with a sufficient radial
component to allow mixing of angular momentum and material. The
instability can roughly be broken into two limits, depending on a critical
baroclinic Richardson number \citep{fuj87},
\be
	Ri_{\rm BC} \equiv 4\lp\frac{R}{H}\rp^2\lp\frac{\Omega}{N}\rp^2
	= 8.5\times10^3\ \mu_{1.33}\temp^{-1}\spin^2.
\ee
When $Ri>Ri_{\rm BC}$, Coriolis effects limit the horizontal scale of
perturbations. This results in two parametrizations for
viscosity estimated from linear theory \citep{fuj93},
\be
       \nu_{\rm BC} = \left\{
              \begin{array}{ccc}
        \displaystyle \frac{ 1}{3}\frac{1}{Ri^{1/2}}H^2\Omega,
                &\hspace{0.2cm}&Ri\le Ri_{\rm BC},\\
         && \\
        \displaystyle \frac{1}{3}\frac{Ri_{\rm BC}}{Ri^{3/2}}H^2\Omega,
                &\hspace{0.2cm}&Ri>Ri_{\rm BC},
              \end{array}
       \right.
       \label{eq:baroclinic}
\ee
where a factor of order unity is usually included in these prescriptions, called
$\alpha_{\rm BC}$, to account for uncertainty
in how linear theory relates to the saturated amplitudes of the
instability. For simplicity, we set $\alpha_{\rm BC}=1$ in our
analysis below.

   By substituting $\nu_{\rm BC}$ into the angular momentum equation
(eq. [\ref{eq:angularmomentum}]), we solve for the shearing profile.
Since we are interested in how the shear relates to the bursting properties,
it useful to write these results in terms of the local accretion rate
$\dot{m}=\dot{M}/(4\pi R^2)$,
which is typically parametrized in terms of the local Eddington rate
\be
	\dot{m}_{\rm Edd} &=& \frac{2m_p c}{(1+X)R\sigma_{\rm Th}}
	\nonumber
	\\
	&=& \frac{1.5\times10^5\ {\rm g\ cm^{-2}\ s^{-1}}}{(1+X)R_6},
	\label{eq:eddington}
\ee
where $X$ is the hydrogen mass fraction and $\sigma_{\rm Th}$ is the
Thomson cross-section. The Richardson number in each case is then
\be
	Ri = \left\{
		\begin{array}{lcc}
	2.2\times10^3\ \mu_{1.33}^{-3/2}\dens\temp^{3/2}\mdot^{-1}\Omega_{0.1},
		&\hspace{0.2cm}&Ri\le Ri_{\rm BC}\\
	&&\\
	4.4\times10^3\ \mu_{1.33}^{-1/4}\dens^{1/2}\temp^{1/4}\mdot^{-1/2}\Omega_{0.1}^{3/2},
		&\hspace{0.2cm}&Ri> Ri_{\rm BC}
		\end{array}
	\right.
	\nonumber
	\\
\ee
where $\dot{m}_{0.1}=\dot{m}/0.1\dot{m}_{\rm Edd}$ and we have used
the hydrogen deficient ($X=0$) value for $\dot{m}_{\rm Edd}$ (eq. [\ref{eq:eddington}]).
The transition to the case $Ri\gtrsim Ri_{\rm BC}$ occurs roughly at depths
$\rho\gtrsim4\times10^6\ {\rm g\ cm^{-3}}\ \temp^{-5/2}$.
The shear rate for each case is
\be
	q_{\rm BC} = \left\{
		\begin{array}{lcc}
	14\ \mu_{1.33}^{5/4}\dens^{-1/2}\temp^{-5/4}\mdot^{1/2}\spin^{-3/2},
		&\hspace{0.2cm}&Ri\le Ri_{\rm BC}\\
		&&\\
	9.9\ \mu_{1.33}^{5/8}\dens^{-1/4}\temp^{-5/8}\mdot^{1/4}\spin^{-7/4},
		&\hspace{0.2cm}&Ri> Ri_{\rm BC}
		\end{array}
	\right.
	\label{eq:qbc}
\ee
This demonstrates that generally $q_{\rm BC}\ll q_{\rm KH}$, so that
the baroclinic instability triggers before the Kelvin-Helmholtz instability.
This prevents the shear rate from ever becoming
large enough for the Kelvin-Helmholtz instability to operate at depths of
$\rho\gtrsim6\times10^4\ {\rm g\ cm^{-3}}$.


\subsection{Other Hydrodynamic Instabilities}

   In addition to the two instabilities described above, there are a number of other
possibilities including, but not limited to, Eddington-Sweet
circulation \citep{von24a,von24b,bk59}, Solberg-H\o iland
instability \citep{was46,tas78,es78},
Goldreich-Schubert-Fricke instability \citep{gs67,fri68}
and Ekman pumping \citep{ped87}. At this time
we avoid assessing each of these individually. As we show below, magnetic fields
are likely to dominate and are interesting since they have received the least attention
in past works.


\section{The Importance of Magnetic Effects}
\label{sec:magnetic}

   Given the larger than order unity shear rates derived above, we estimate the
consequences this has for a magnetic field.
The point we want to emphasize is that even a reasonably
small field will be wrapped by the
shear flow until it becomes dynamically important.

   Assuming shellular rotation, a component of radial
field is stretched to have a toroidal component, $B_\phi = n B_r$, where $n$ is
the number of differential turns, given as $n=q\Omega t$, and $t$ is the duration
of the shearing. The toroidal field growth is very fast. For the Kelvin-Helmholtz
case, $B_\phi\sim B_r$ in merely $\sim10^{-6}\ {\rm s}$ ($\sim10^{-4}\ {\rm s}$ for
the baroclinic case). As the toroidal field becomes larger it exerts an azimuthal
stress on the shearing layer equal to $B_rB_\phi/4\pi$, which can be written
as an effective viscosity, $\nu_e$,
\be
	\frac{B_rB_\phi}{4\pi}=\rho\nu_e q\Omega.
	\label{eq:veff}
\ee
In a timescale $t\approx H_\Omega^2/\nu_e$ this torquing significantly
decreases the spin of the layer, where
$H_\Omega=dz/d\ln\Omega=R/q$ is the shearing scale height.
Setting $B_\phi=q\Omega tB_r$, we solve for the
critical initial radial field needed to affect the shearing,
\be
	B_{r,\rm crit} = (4\pi \rho)^{1/2}\frac{H_\Omega}{t} = (4\pi \rho)^{1/2}\frac{R}{qt},
	\label{eq:bcrit}
\ee
so that $t$ is basically the Alfv\'{e}n travel time through a shearing scale height.
Using $\rho\approx10^6\ {\rm g\ cm^{-3}}$ as burning density
and $t\approx 1\ {\rm hr}$, a fiducial timescale for accumulation,
our estimates for $q_{\rm KH}$ and $q_{\rm BC}$ imply
$B_{r,\rm crit}\sim10^{4}\ {\rm G}$ and $\sim10^5\ {\rm G}$, respectively
Any initial field larger than this gets so wound up that Lorentz
forces alter the shearing profile.

   It is possible that the intrinsic magnetic field of the NS may be large enough
that magnetic stresses never allow the shear rates to become so large.
\citet{spr99} argued that this depends
on the ``rotational smoothing time,'' the timescale for non-axisymmetric components
of the magnetic field to be expelled as differential rotational brings field with opposite
polarities together. This timescale is estimated as
\be
	t_\Omega = \lp\frac{3R^2\pi^2}{\eta\Omega^2q^2}\rp^{1/3},
\ee
where $\eta$ is the magnetic diffusivity, which in the
non-degenerate limit is given by \citep{spi62}
\be
	\eta \approx 0.7\ {\rm cm^2\ s^{-1}}T_8^{-3/2}.
\ee
If $t_\Omega$ is larger than the Alfv\'{e}n travel time through the layer (given
by $t$ in eq. [\ref{eq:bcrit}]), then there is sufficient time for magnetic torques
to act and the shearing is merely a perturbation on the magnetic field that is
quickly damped away. On the other hand, if $t_\Omega$ is sufficiently small
then the shear dominates and the magnetic field is made axisymmetric
on a timescale $t_\Omega$. Substituting $t_\Omega$ into equation (\ref{eq:bcrit}),
we estimate the critical field, below which shear dominates,
\be
	B_\Omega &=& (4\pi\rho)^{1/2}\lp\frac{\eta\Omega^2R}{3\pi^2}\rp^{1/3}q^{-1/3}
	\nonumber
	\\
	&=&1.3\times10^7\ {\rm G}\ \dens^{1/2}\temp^{-1/2}
	\spin^{2/3}q^{-1/3}.
\ee
The lack of persistent pulsations from accreting NSs implies a dipole field
strength $\lesssim5\times10^7\ {\rm G}$ \citep{pb05}.
We therefore consider it plausible that they
may have intrinsic magnetic fields $<B_\Omega$. In this case, we
expect a very axisymmetric magnetic field to be created in a timescale of
$t_\Omega\sim200\ {\rm s}$,
which is then wrapped until it becomes dynamically important.


\subsection{The Tayler-Spruit Dynamo}
\label{sec:tssummary}

   As the toroidal magnetic field continues to wrap, it becomes increasingly important
to the dynamics of the shearing and is also subject to magnetohydrodynamic
instabilities. The combination of these effects have been shown to give rise to the Tayler-Spruit
dynamo \citep{spr02}. In this picture, shearing grows the toroidal field, which
then initiates Tayler instabilities \citep[non-axisymmetric, pinch-like instabilities
including stratification,][]{tay73,spr99}. This turbulently creates poloidal field
components that once again shear to be toroidal. This cycle continues, creating a steady-state
field.

   The minimum shear needed for this process to operate can be argued
simply. (See Spruit 2006 for a mathematical derivation that uses the dispersion
relation from Acheson 1978.) We note that \citet{dp07} have recently given an
alternate prescription for this same mechanism using solely heuristic arguments.
Since their results have not been shown consistent with a more rigorous mathematical
analysis we consider Spruit's conclusions more reliable at this time.
A vertical perturbation, $l_z$, is limited by buoyancy forces
to be (eq. [6] from Spruit 2002)
\be
	l_z<R\wa/N,
	\label{eq:lmax}
\ee
where $\wa=B/[(4\pi\rho)^{1/2}R]$ is the Alfv\'{e}n frequency. At small
lengthscales, magnetic diffusion damps out perturbations. In the
limit of $\Omega\gg\wa$, which we are considering, the Tayler instability
growth rate is $\sigma_B=\wa^2/\Omega$, so that
\be
	l_z^2>\eta/\sigma_B = \eta\Omega/\wa^2.
	\label{eq:lmin}
\ee
Combining these two relations gives the minimum
$\wa$ needed for the dynamo to act,
\be
	\lp\frac{\wa}{\Omega}\rp^4 > \frac{\eta}{R^2\Omega}\lp\frac{N}{\Omega}\rp^2.
	\label{eq:limit}
\ee
During the timescale for Tayler instability, $\sigma_B^{-1}$, $B_r$ is stretched
into $B_\phi$ by an amount
\be
	B_\phi=\sigma_B^{-1}q\Omega B_r.
	\label{eq:growth}
\ee
The largest amplification is achieved for magnetic fields that extend the largest
radial lengthscale available, so that assuming equation (\ref{eq:lmax}) is marginally
satisfied along with the induction equation we find $B_r/B_\phi=l_z/R=\wa/N$.
Combining this with equation (\ref{eq:growth})
we obtain $q=(N/\Omega)(\wa/\Omega)$. Substituting this into equation (\ref{eq:limit}),
we find \citep{spr02}
\be
	q_{\rm min}=\lp\frac{N}{\Omega}\rp^{7/4}\lp\frac{\eta}{R^2N}\rp^{1/4}
	\label{eq:qmin}
\ee
as the minimum shear needed for the Tayler instability to operate. Though this result
is consistent with more rigorous analysis \citep{spr06}, it should
be viewed with some caution as we apply it to accreting NSs. The thin shell
geometry we consider is quite different than the spherical geometries typically
used when invoking the Tayler-Spruit dynamo. Simulations by \citet{bra06} demonstrate
that the Tayler instability is strongest along the rotation axis, which is only realized
at the poles in the NS case. Since we find the dynamo to be so much stronger than
any hydrodynamic transport mechanisms, we consider it to be a reasonable
approximation for these magnetic effects, even if its strength is decreased due
to geometry.

  The value of $N$ used to evaluate equation (\ref{eq:qmin}) depends
on what is supplying the buoyancy.
We follow \citet{spr02} and separately consider cases of
$N_\mu\gg N_T$ and $N_\mu\ll N_T$, denoted as case 0 and 1, respectively.
In case 1 non-adiabatic effects become important when $\eta/K<1$, and
we must take $N_T^2\rightarrow (\eta/K)N_T^2$
(analogous to the above analysis of the secular shear instability). For this case,
which dominates for most of the envelope, we find
\be
	q_{\rm min}=0.10\ \opac^{3/4}\dens^{3/2}\temp^{-9/2}\spin^{-7/4}.
\ee
Both $q_{\rm KH}$ and $q_{\rm BC}$ are considerably above this,
thus the Tayler-Spruit dynamo activates long before the onset of purely
hydrodynamic instabilities.

   By assuming that equation (\ref{eq:limit}) is marginally satisfied \citet{spr02}
derived the steady-state field strengths. We summarize the
relevant prescriptions needed for our study. The steady-state azimuthal
and radial field components are
 \be
 	B_{\phi0} &=& (4\pi\rho)^{1/2}Rq\Omega^2/N_\mu,
	\\
	B_{r0} &=& q\lp\frac{\Omega}{N_\mu}\rp^2B_{\phi0},
\ee
for case 0. For case 1,
\be
	B_{\phi1} &=& (4\pi\rho)^{1/2}R\Omega q^{1/2}
		\lp\frac{\Omega}{N_T}\rp^{1/8}\lp\frac{K}{R^2N_T}\rp^{1/8},
	\\
	B_{r1} &=& \lp\frac{\Omega}{N_T}\rp^{1/4}\lp\frac{K}{R^2N_T}\rp^{1/4}B_{\phi1}.
\ee
The effective viscosities, as defined by equation (\ref{eq:veff}), are
\be
	\nu_0 &=& R^2\Omega q^2\lp\frac{\Omega}{N_\mu}\rp^4,
	\\
	\nu_1 &= & R^2\Omega\lp\frac{\Omega}{N_T}\rp^{1/2} \lp\frac{K}{R^2N_T}\rp^{1/2}.
\ee
Although these viscosities are appropriate for angular momentum transport, mixing
of material is less efficient since it requires expending work to exchange fluid
elements (versus just exerting shear stresses). The mixing diffusivities are
\be
	D_0 &=& R^2\Omega q^4\lp\frac{\Omega}{N_\mu}\rp^6,
	\label{eq:d0}
	\\
	D_1 &=& R^2\Omega q\lp\frac{\Omega}{N_T}\rp^{3/4}\lp\frac{K}{R^2N_T}\rp^{3/4}.
	\label{eq:d1}
\ee
which is just equal to the effective turbulent magnetic diffusivity. In the Appendix we show that these
prescriptions are consistent with energy conservation considerations.

\subsection{Shearing Profile}
\label{sec:shearing}

   The Tayler-Spruit dynamo transports angular momentum, causes viscous heating,
and mixes material. We follow the procedure we used for the baroclinic
case and assume steady-state angular momentum transport. This is a good
approximation when the viscous timescale, $t_{\rm visc}=H^2/\nu$, is less than
the timescale of accretion. For these
initial estimates we focus on the $N_T\gg N_\mu$ limit (case 1) since this
dominates except at compositional boundaries (which we revisit in \S \ref{sec:nomix}).
Substituting $\nu_1$ into equation (\ref{eq:angularmomentum}), the shear rate
for the Tayler-Spruit dynamo is
\be
	q_{\rm TS} = \frac{\dot{m}}{R\rho\Omega}\frac{\Omega_{\rm K}}{\Omega}
		\lp\frac{N}{\Omega}\rp^{1/2}\lp\frac{R^2N}{K} \rp^{1/2}.
	\label{eq:qts1}
\ee
Scaling to values appropriate for accreting NSs,
\be
	q_{\rm TS} &=& 0.38\ \opac^{1/2}\temp^{-2}
		\mdot\spin^{-5/2}.
	\label{eq:qts}
\ee
Note the scalings with $\dot{m}$ and $\Omega$. At high $\dot{m}$ angular
momentum is fed into the star faster, creating more shear. At low
$\Omega$ the shear is greater because of the larger relative angular
speed between the accreted material and the NS. These are generic feature
we expect for any viscous mechanism (compare the scaling of $q_{\rm TS}$
with $q_{\rm KH}$ and $q_{\rm BC}$ from eqs.
[\ref{eq:qkh}] and [\ref{eq:qbc}], respectively).
The viscosity that gives the smallest shear rate will likely be the most important
at a given depth. Using this criterion, we find that the Tayler-Spruit dynamo is dominant for
densities $\rho\gtrsim3\ {\rm g\ cm^{-3}}\ \temp^{-5/2}\mdot^2\spin^{-3}$,
or $\rho\gtrsim5\times10^3\ {\rm g\ cm^{-3}}$ for $T\approx5\times10^6\ {\rm K}$.
At shallower depths, Kelvin-Helmholtz instabilities damp the shear, which is consistent
with the use of hydrodynamic instabilities by \citet{is99} for understanding the initial
spreading of accreted material in the boundary layer.

   The steady-state magnetic field components are
\be
	B_\phi &=& 1.3\times10^{10}{\rm\ G}\ \opac^{-1/8}\dens^{1/4}\temp^{-1/2}
		\mdot^{1/2}\spin^{-1/8},
	\\
	B_r &=& 2.1\times10^5{\rm\ G}\ \opac^{-3/8}\dens^{-1/4}\temp^{1/2}
		\mdot^{1/2}\spin^{1/8}.
\ee
\citet{cb00} argue that if $B_r\gtrsim10^6\ {\rm G}$ it would
become dynamically important to determining the drift of X-ray burst oscillations \citep{mun02}.
The interaction of such fields with the shearing of an X-ray burst from an accreting
millisecond pulsar has been explored by \citet{lov07}.
At high accretion rates, $B_r$ increases and comes close to this limit, suggesting
that these magnetic fields may be important for understanding the dynamics
of X-ray burst oscillations, even from NSs that do not show persistent pulsations
\citep[which are preferentially seen at high accretion rates,][]{mun04}.

\subsection{Viscous Heating}
\label{sec:heating}

   Viscous shearing heats the surface layer, which can also be thought
of as the rate of magnetic energy destruction as the dynamo
builds and destroys magnetic field \citep{mm04}. The heating rate per unit mass is
\be
	\epsilon = \frac{1}{2}\nu\lp q\Omega\rp^2,
\ee
which for the Tayler-Spruit dynamo becomes
\be
	\epsilon_{\rm TS} &=& 5.6\times10^{10}\ {\rm ergs\ g^{-1}\ s^{-1}}
		\opac^{1/2}\dens^{-1}\temp^{-2}
		\mdot^2\spin^{-3/2}.
		\nonumber
		\\
\ee
To put this number into perspective we express it
in terms of the energy released per accreted nucleon. This
is found by multiplying the above result by the total mass per unit
area down to depth of interest, $y\approx H\rho$ (the column depth),
and dividing by $\dot{m}$. We then find
\be
	\frac{dE_{\rm TS}}{d\ln y} &=& 0.13\ {\rm keV\ nuc^{-1}}
		\mu_{1.33}^{-1}\opac^{1/2}\temp^{-1}\mdot\spin^{-3/2}.
\ee
In comparison, the thermal energy per nucleon
at $10^8\ {\rm K}$ is $\approx~10\ {\rm keV\ nuc^{-1}}$
and burning of helium into carbon releases
$\approx~0.6\ {\rm MeV\ nuc^{-1}}$. We conclude that viscous
processes do not heat the layer sufficiently to alter X-ray bursts.

\subsection{Turbulent Mixing}
\label{sec:mixing}

   It is not immediately clear how much of an observational impact is provided by
the magnetic fields and viscous heating estimated
above. In contrast, as we shall show,
shear mixing produced by the dynamo could have important ramifications
for the structure and composition of the surface layers. For this reason, we devote
most of the remainder of our study to investigating the consequences of mixing.

   We first consider some estimates that highlight mixing's
importance. For this, we parametrize the mixing diffusivity as $\alpha_{\rm TS} D$,
where $D$ is given by equations (\ref{eq:d0}) or (\ref{eq:d1}) and
$\alpha_{\rm TS}$ is a factor of order unity that accounts for uncertainties in the
Tayler-Spruit prescription. The features we find are
general enough that other viscosities can be incorporated by increasing
or decreasing $\alpha_{\rm TS}$.

   The material becomes fully mixed over a scale height, $H$, in a time
$\tmix=H^2/(\alpha_{\rm TS} D)$, which gives for case 1 ($N_T\gg N_\mu$),
\be
	\tmix &=& 4.3\times10^2\ {\rm s}\ \alpha_{\rm TS}^{-1}
		\mu_{1.33}^{-2}\opac^{1/4}\dens^{3/2}\temp
		\mdot^{-1}\spin^{3/4},
	\label{eq:tmix}
\ee
Accretion is also advecting material downward, which
happens in a timescale $\tacc=y/\dot{m}\approx\rho H/\dot{m}$,
\be
	\tacc = 2.2\times10^3\ {\rm s}\ \mu_{1.33}^{-1}\dens\temp\mdot^{-1}.
	\label{eq:tacc}
\ee
The ratio of these timescales is
\be
	\frac{\tmix}{\tacc} &=& 0.19\ \alpha_{\rm TS}^{-1}\mu_{1.33}^{-1}\opac^{1/4}
		\dens^{1/2}\spin^{3/4}.
	\label{eq:timeratio}
\ee
The scaling with $\rho$ shows that mixing is generally more important than
advection at shallow depths ($\rho\lesssim3\times10^7\ {\rm g\ cm^{-3}}$),
while the scaling with $\Omega$ shows
that mixing is more important for slower spinning NSs (as expected from
our discussion of shear rates in \S \ref{sec:shearing}).


\section{Accumulating, Non-mixed Models}
\label{sec:nomix}

   We now use numerical calculations to consider the angular
momentum transport through the surface layers. In this section we calculate
accumulating models without directly incorporating the effects from mixing.
This verifies many of the analytic estimates derived above and motivates
when mixing must be included (which is done for models in \S \ref{sec:mix}).


\subsection{Shear Profile Calculations}
\label{sec:shearprofile}

   We calculate the envelope profile for helium accumulating on an iron ocean
(which represents the iron-peak ashes from previous X-ray bursts). As
discussed in \S \ref{sec:baroclinic}, we approximate the surface
as having a constant gravitational acceleration, and
plane-parallel geometry. In addition to using $z$ as our radial
coordinate, we find it useful to use the column
depth, $y$, defined as $dy=-\rho dz$, giving a pressure $P=gy$ from hydrostatic
balance. We solve for $\rho$ using the analytic equation
of state from \citet{pac83}. For the liquid phase, when $1\le\Gamma\le173$
where $\Gamma=(4\pi n_i/3)^{1/3}Z^2e^2/(k_{\rm B}T)$, $Z$ is the charge per
ion, and $n_i$ is the ion density, we include the ionic free energy of \citet{cp98}.

   During the accumulating phase, a negligible amount of helium burning
takes place, so we assume that the flux is constant and set by heating from
the crust. Previous studies of accreting NSs have shown that the interior
thermal balance is set by electron captures, neutron emissions, and
pycnonuclear reaction in the inner crust \citep{mph90,zdn92,bb97,bb98,bro00,bro04}
which release $\approx1\textrm{ MeV}/m_p\approx10^{18}\ {\rm ergs\ g}^{-1}$
\citep{hz90,hz03}. Depending on the accretion rate and thermal structure of
the crust, this energy is either conducted into the core or released into
the ocean such that for an Eddington accretion rate up to $\approx92\%$
of the energy is lost to the core and exits as neutrinos \citep{bro00}. We
therefore set the heating to $150\ {\rm keV\ nuc^{-1}}$, giving a flux
$2.2\times10^{21}\ {\rm ergs\ cm^{-2}\ s^{-1}}\langle\dot{m}\rangle_{0.1}$,
where $\langle\dot{m}\rangle_{0.1}$ is the time average accretion rate in
units of $0.1\dot{m}_{\rm Edd}$ and the average is over timescales of order
the thermal time of the crust (many years). For simplicity we assume
$\langle\dot{m}\rangle=\dot{m}$. Recent calculations by \citet{gup06}
suggest that heating is stronger than previously thought, but
not sufficiently high to qualitatively change our results. We ignore
the additional flux from compressional heating because it only
contributes $\sim c_pT\sim10\ {\rm keV\ nuc^{-1}}$ \citep{bil98a}.

   In a one-zone estimate ignition occurs at the base of the helium layer
when
\be
	\frac{d\epsilon_{3\alpha}}{dT} = \frac{d\epsilon_{\rm cool}}{dT},
	\label{eq:stability}
\ee
\citep{fhm81}
where $\epsilon_{3\alpha}$ is the heating rate from triple-$\alpha$ reactions
(for which we use the rate from Fushiki \& Lamb 1987 for numerical calculations),
\be
	\epsilon_{\rm cool}=\frac{4\sigma_{\rm SB}T^4}{3\kappa y^2},
\ee
and the derivatives are
both taken at constant pressure. We consider models where the column of accreted
helium $y_{\rm acc}$ that is just at this ignition point. This allows the
maximum amount of time for angular momentum redistribution during the
accumulating phase. The He/Fe boundary is assumed sharp,
since the timescale for diffusion between the layers is much longer
than the timescale of accumulation \citep{bbc02}.

   We solve for the temperature profile by integrating the radiative
diffusion equation,
\begin{eqnarray}
        F = \frac{16\sigma_{\rm SB}T^3}{3\kappa}\frac{dT}{dy}.
        \label{eq:flux}
\end{eqnarray}
The opacity is set using electron-scattering \citep{pac83},
free-free, and conductive opacities \citep{sch99}.
There is a sharp change in opacity at the He/Fe boundary
(because the high $Z$ of the iron makes it more opaque),
which means that the top of the iron layer is convectively unstable.
The characteristic convective velocity estimated from $F\approx\rho V_{\rm conv}^3$
is $V_{\rm conv}\sim10^5\ {\rm cm\ s^{-1}}$, which is much less
than the local sound speed of $\sim10^8\ {\rm cm\ s^{-1}}$. This means
that the convection is very efficient, and we therefore simply set
$(d\ln T/d\ln P)_*=\nabla_{\rm ad}$ in the convective region. This
convective region has not been noted in previous studies, though
a super-adiabatic temperature gradient is apparent in the figures
shown in both \citet{bbc02} and \citet{cum03}. It is not yet clear
how this impacts the bursting properties of NSs,
and we delay exploring this in detail for a future study. Nevertheless,
we must include convection so as to have accurate
differentially rotating profiles.

\begin{figure}
\epsscale{1.3} 
\plotone{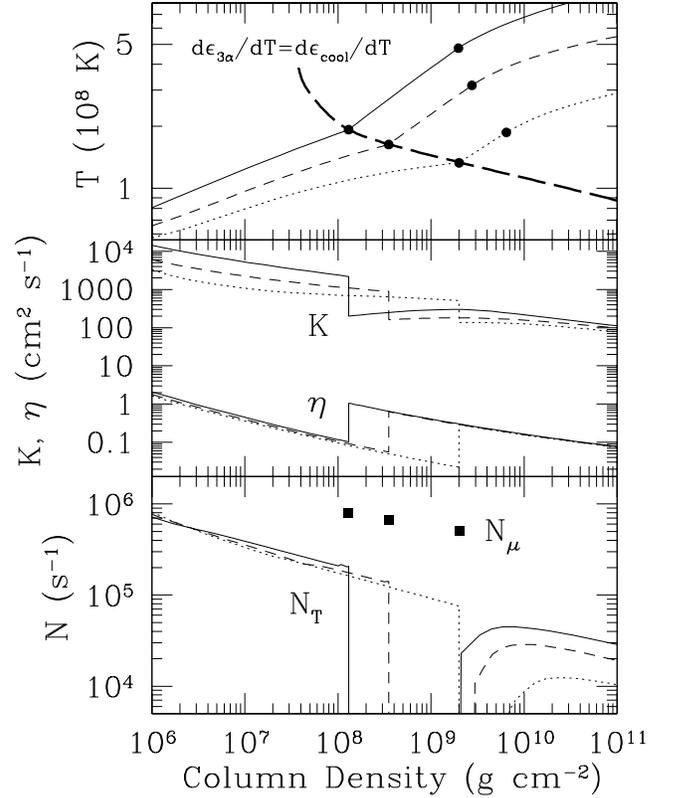}
\caption{Accumulating models for accretion rates of
$0.1\dot{m}_{\rm Edd}$ ({\it dotted line}), $0.3\dot{m}_{\rm Edd}$
({\it short-dashed line}), and $1.0\dot{m}_{\rm Edd}$ ({\it solid line}).
For each model the base of the helium layer is taken to be at the unstable
triple-$\alpha$ ignition depth where $d\epsilon_{3\alpha}/dT=d\epsilon_{\rm cool}/dT$
({\it thick long-dashed line} in top panel). The top panel plots the temperature,
and the solid circles mark the top and bottom of the convective zone.
The middle panel plots the thermal diffusivity, $K$,
and magnetic diffusivity, $\eta$. The bottom panel shows the Brunt-V\"{a}is\"{a}l\"{a}
frequency $N$ for both thermal ($N_T$, {\it lines}) and compositional ($N_\mu$,
{\it solid squares}) contributions.}
\label{fig:accumulating}
\epsscale{1.0}
\end{figure}
   In Figure \ref{fig:accumulating} we plot the temperature profile, $K$ and $\eta$,
and $N$ for three accumulating models. The magnetic diffusivity is set using
the conductivity from \citet{sch99}. The base of the accumulating helium is set
where unstable helium ignition occurs, as designated by the thick long-dashed
line in the top panel (eq. [\ref{eq:stability}]). The convective zone begins just
below this and is bracketed by solid circles at its top and bottom. The convective
zone is also seen in the plot of $N$ since $N_T$ is effectively zero in this
region. The change in composition at the He/Fe boundary gives a large
buoyancy contribution, which we estimate as
\be
	N_\mu\approx \lp\frac{g}{H}\Delta\ln\mu\rp^{1/2},
	\label{eq:bruntmu}
\ee
where $\Delta\ln\mu\approx0.44$ is the logarithmic change in the mean
molecular weight at the boundary. We denote this by a solid square for each
model in the bottom panel. For the majority of the profile the Tayler-Spruit
dynamo is given by case 1 ($N_T\gg N_\mu$). Since
$K\gg\eta$, the perturbations creating the dynamo are non-adiabatic for
case 1, as we described in \S \ref{sec:tssummary}. The only place case 0
($N_\mu\gg N_T$) is important is at the He/Fe boundary.

   We next solve for the shear rates by solving equation (\ref{eq:angularmomentum})
with either $\nu_0$ or $\nu_1$, depending on which case is appropriate.
We assume angular momentum transport does not
affect the thermal or compositional structure, in
other words, that the transport is just happening ``in the background.''
This allows us to assess which effects are crucial for subsequent iterations that
include angular momentum transport in the actual structure
calculation.

\begin{figure}
\epsscale{1.25} 
\plotone{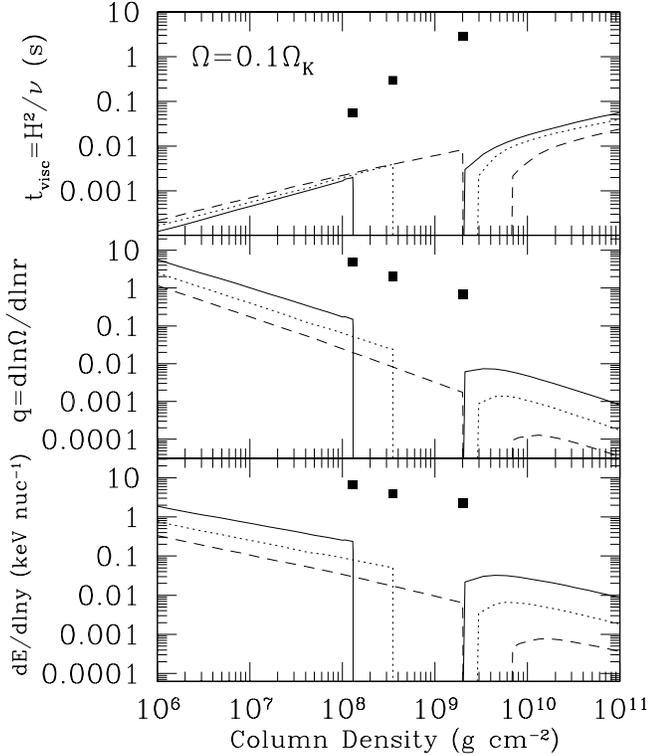}
\caption{Angular momentum transport in the NS surface layers, for a spin of
$\Omega=0.1\Omega_{\rm K}$, and accretion rates of
$0.1\dot{m}_{\rm Edd}$ ({\it dotted line}), $0.3\dot{m}_{\rm Edd}$
({\it dashed line}), and $1.0\dot{m}_{\rm Edd}$ ({\it solid line}).
The panels display ({\it from top to bottom}) the viscous
timescale for angular momentum
transport across a scale height, $t_{\rm visc}$, the shear rate, $q$, and the
viscous energy deposition per logarithm column, $dE/d\ln y$.
The squares indicate the corresponding values due to the compositional
discontinuity at the base of the accumulating layer.}
\label{fig:transport1}
\epsscale{1.0}
\end{figure}
   In Figure \ref{fig:transport1} lines denoted the profiles calculated assuming
case 1 of the Tayler-Spruit dynamo. We assume that
the convection is effectively instantaneous in transporting material and
angular momentum, thus all the quantities become very small in the convective region. This
is appropriate since the convective overturn timescale,
$H/V_{\rm conv}\sim10^{-4}\ {\rm s}$, is much less than the timescales for
mixing or accretion. The squares indicate the corresponding values due to the
compositional discontinuity found by using case 0. Since the viscous timescale,
$t_{\rm visc}=H^2/\nu$, is much less than
the timescale it took to accrete to this base column
$\tacc=y_{\rm acc}/\dot{m}\sim10^3-10^5\ {\rm s}$ (for the range of $\dot{m}$
considered) our steady-state assumption is valid.

   These calculations highlight
the importance of the compositional jump since $t_{\rm visc}$, $q$, and
$dE/d\ln y$ are all amplified here. This is because the large buoyancy reduces
angular momentum transport across this boundary.
The viscous time is nearly independent of accretion rate in regions where
$N_T\gg N_\mu$ because the viscosity in case 1 is independent of $q$. The energy
deposition is always much smaller than the heat coming the crust, and
it falls off somewhat faster with depth than expected from the estimate presented in
\S \ref{sec:heating} due to electron degeneracy effects decreasing $N_T$.
New heat sources at a depth of $\approx10^{12}\ {\rm g\ cm^{-2}}$
could ease the difficulty calculations have in recreating
the low ignition columns needed to explain superburst recurrence times
\citep{cum06}, but this viscous heating is not nearly enough to correct this problem.

\begin{figure}
\epsscale{1.25} 
\plotone{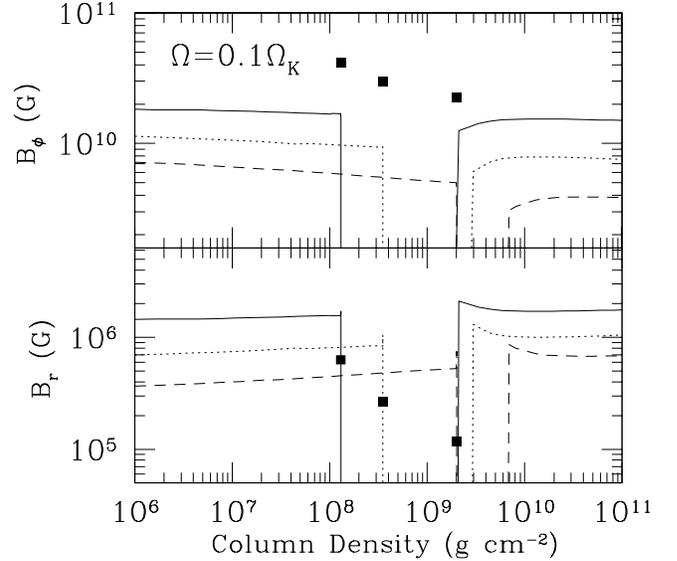}
\caption{Azimuthal and radial field components for a spin of $\Omega=0.1\Omega_{\rm K}$.
Lines have the same meaning as in Fig. \ref{fig:transport1}.}
\label{fig:bfields}
\epsscale{1.0}
\end{figure}

   In Figure \ref{fig:bfields} we plot the magnetic fields found within the radiative
zones for the models from Figure \ref{fig:transport1}. We assume that within the
convective zone the dynamo is not able to operate, and do not calculate
a magnetic field here. It is also interesting
to compare these fields to those derived by \citet{czb01}, who calculated the
steady-state fields expected when Ohmic diffusion balances advection through accretion.
Their principal result was that the steady-state horizontal field drops by
$\approx\dot{m}/0.02\dot{m}_{\rm Edd}$ orders of magnitude from the crust up through the ocean.
In contrast, the magnetic fields we find are nearly constant with depth.
We therefore do not expect these fields to persist if the
accretion ceases for a time and instead to be expelled on an Ohmic diffusion time ($\sim\ {\rm days}$
near the top of the ocean). When the dynamo is active, the steady-state is reached quickly
enough that Ohmic diffusion can be ignored.

\begin{figure}
\epsscale{1.25} 
\plotone{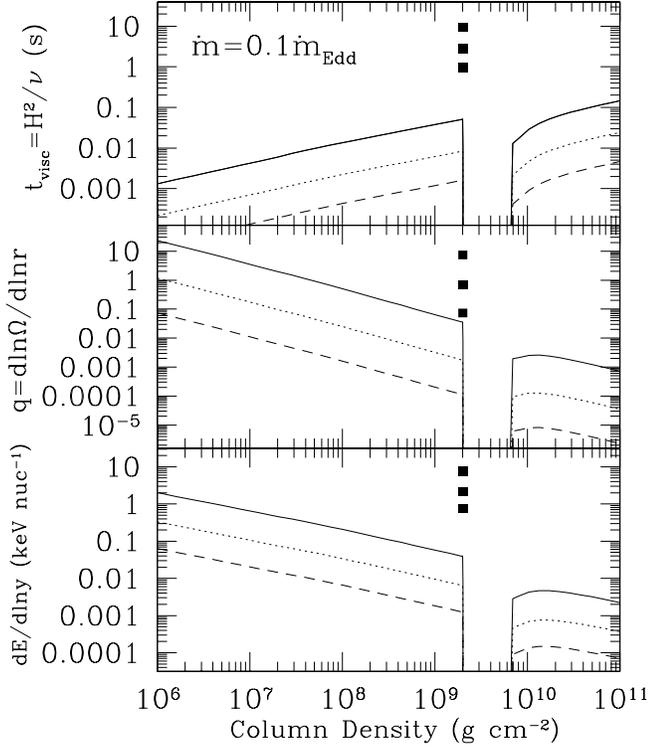}
\caption{Same as Fig. \ref{fig:transport1}, but for spins of $0.03\Omega_{\rm K}$ ({\it solid line}),
$0.1\wk$ ({\it dotted line}), and $0.3\Omega_{\rm K}$ ({\it dashed line}), all for
$\dot{m}=0.1\dot{m}_{\rm Edd}$.}
\label{fig:transport2}
\epsscale{1.0}
\end{figure}

   In Figure \ref{fig:transport2} we compare what happens as the spin is changed by plotting
$\Omega=0.03$, 0.1, and $0.3\Omega_{\rm K}$ ($67$, $220$, and $670\ {\rm Hz}$),
all for $\dot{m}=0.1\dot{m}_{\rm Edd}$. Note that
the shearing is most dramatic at smaller $\Omega$. In fact, the shearing and heating
profiles are very sensitive to the value of $\Omega$, as was demonstrated by the
analytic estimates. We do not plot the associated magnetic field for these models since
we have already plotted some examples in Figure \ref{fig:bfields} and the analysis of
\S \ref{sec:shearing} provides adequate estimates.

\begin{figure}
\epsscale{1.2} 
\plotone{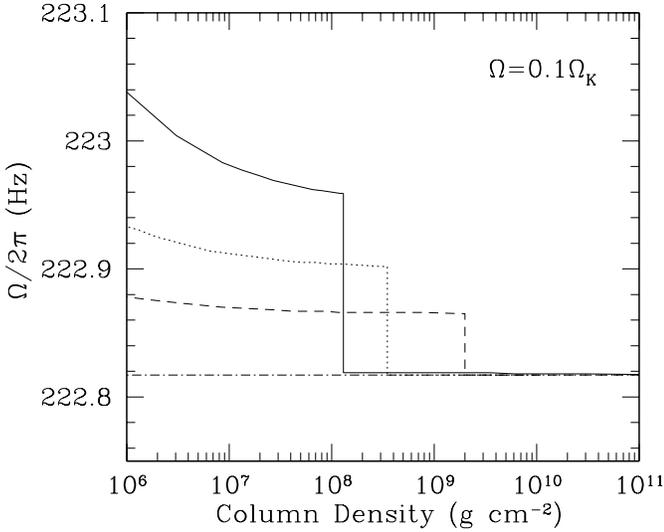}
\caption{The spin frequency as a function of column depth for $\Omega=0.1\Omega_{\rm K}$,
with $\dot{m}=0.1$ ({\it dashed line}), 0.3 ({\it dotted line}),
and $1.0\dot{m}_{\rm Edd}$ ({\it solid line}). The dot-dashed
line shows a constant spin frequency for comparison.}
\label{fig:spin}
\epsscale{1.0}
\end{figure}

   These plots of $q$ show that very little shearing is present.
To emphasize this fact, in Figure \ref{fig:spin} we plot the actual spin frequency
as a function of depth found by integrating $q$ for a range of accretion rates. The
discontinuity at the He/Fe boundary complicates this estimate. Noting that
\be
	\Omega = \int q\Omega d\ln r = - \int q\Omega\frac{H}{R}d\ln y,
\ee
we approximate the spin change at this boundary as
\be
	\Delta\Omega\approx q\Omega H/R.
\ee
The change of spin is generally $\lesssim0.5\ {\rm Hz}$ across the accumulating layer, with
the majority of the spin change occurring at the compositional boundary. The layer
is very nearly in uniform rotation.


\subsection{Mixing and the Compositional Barrier}

   The above calculations confirm that the viscosity is
too large for either appreciable shearing or viscous heating.
As demonstrated in \S \ref{sec:mixing}, mixing should be important,
but Figures \ref{fig:transport1} and \ref{fig:transport2} argue that
we must take into account the large $N_\mu$ at the He/Fe boundary.
Assuming that $N_\mu$
scales like equation (\ref{eq:bruntmu}), we estimate
\be
	N_\mu=1.6\times10^{6}\ {\rm s^{-1}}\ \mu_{1.33}^{1/2}\temp^{-1/2}(\Delta\ln\mu/0.44)^{1/2}.
\ee
We substitute this into $\nu_0$ to solve for $q$ using equation
(\ref{eq:angularmomentum}), which is used to
estimate a mixing timescale at the boundary,
\be
	t_{\rm mix} &=&H^2/(\alpha_{\rm TS} D_0)
	\nonumber
	\\
	&=&1.7\times10^3{\rm s}\ \alpha_{\rm TS}^{-1}\mu_{1.33}^{-5/3}\dens^{4/3}\temp^{5/3}
		\mdot^{-4/3}\spin\lp\frac{\Delta\ln\mu}{0.44}\rp^{1/3}.
		\nonumber
		\\
\ee
Using equation (\ref{eq:tacc}) we find a ratio of
\be
	\frac{\tmix}{\tacc}\approx0.77\ \alpha_{\rm TS}^{-1}\mu_{1.33}^{-2/3}
		\dens^{1/3}\temp^{2/3}\mdot^{-1/3}\spin\lp\frac{\Delta\ln\mu}{0.44}\rp^{1/3}.
		\nonumber
		\\
\ee
Unlike in equation (\ref{eq:timeratio}), this new ratio depends on $\dot{m}$.
This is because the viscosity in case 0 has a different dependence on
$q$. Now as $\dot{m}$ is increased
$\tmix$ decreases faster than $\tacc$. Above a critical accretion rate of
\be
	\dot{m}_{\rm crit,1}=4.6\times10^{-2}\ \dot{m}_{\rm Edd}\ \alpha_{\rm TS}^{-3}
		\mu_{1.33}^{-2}\dens\temp^2\spin^3\lp\frac{\Delta\ln\mu}{0.44}\rp,
		\nonumber
		\\
	\label{eq:mdotcrit1}
\ee
mixing can no longer be ignored. For densities and temperatures expected
at the base of the accumulating layer, this critical accretion rate
lies in the range of $0.1-1.0\dot{m}_{\rm Edd}$. Note that this depends
very strongly on the spin rate, $\dot{m}_{\rm crit,1}\propto\Omega^3$,
thus we expect the slower spinning NSs to be considerably more affected by mixing.

\begin{figure}
\epsscale{1.3} 
\plotone{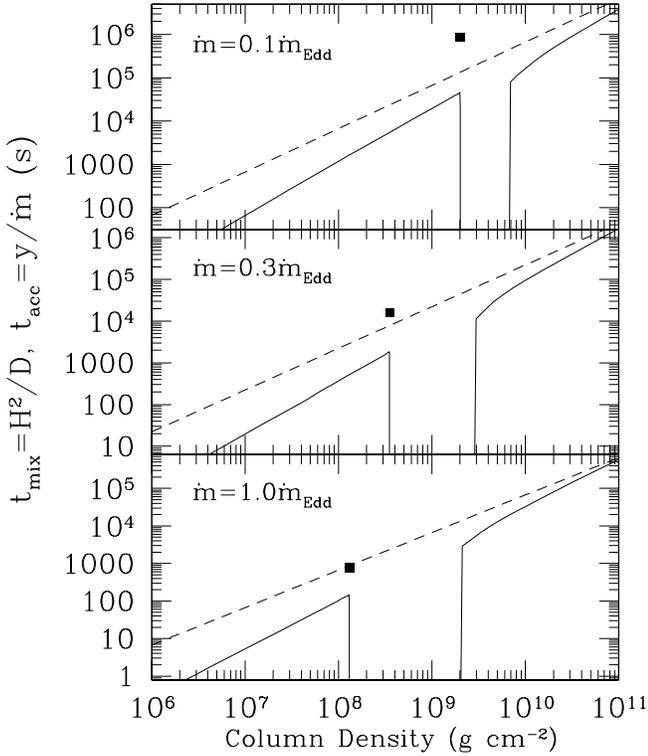}
\caption{Comparison of the mixing timescale,
$\tmix$ ({\it solid line}) versus the accretion timescale
$\tacc$ ({\it dashed line}) for $\dot{m}=0.1, 0.3,$ and $1.0\dot{m}_{\rm Edd}$
({\it top to bottom panel}; all with $\Omega=0.1\Omega_{\rm K}$ and $\alpha_{\rm TS}=1$). The solid
square denotes $\tmix$ due to the compositional discontinuity.
At low $\dot{m}$, the squares are above the dashed line,
demonstrating that $\tmix>\tacc$ at the base of the accumulating
layer, which prevents mixing to larger depths.
At sufficiently high $\dot{m}$, $\tmix<\tacc$ at depths below
the accumulation depth, so that mixing
between helium and iron can occur.}
\label{fig:transport3}
\epsscale{1.0}
\end{figure}
   To test these analytic estimates we compare the mixing and accretion timescales in
Figure \ref{fig:transport3}. From the top panel to the bottom panel we increase
$\dot{m}$ (fixing $\Omega=0.1\wk$). At low $\dot{m}$, the He/Fe boundary
at $y_{\rm acc}$ (shown by the filled squares)
acts as a barrier to mixing since $\tmix>\tacc$ at this
depth. When this happens it is a good approximation
to ignore mixing and assume two separate layers during the accumulation
phase. As $\dot{m}$ increases, $\tmix$ at $y_{\rm acc}$
becomes less and less until finally $\tmix<\tacc$, so that material should be mixed
past $y_{\rm acc}$. When this occurs, our accumulating model can no longer ignore
mixing. The mixing between helium
and iron occurs down to a depth where $\tmix$ is equal to the length of time
accretion has been taking place, $\tacc=y_{\rm acc}/\dot{m}$. The
key point we want to emphasize is that {\it because of the buoyancy
barrier, the effect of mixing turns on abruptly, and when it does,
mixing will occur well past} $y_{\rm acc}$.


\section{The Effects of Turbulent Mixing}
\label{sec:mix}

\begin{figure}
\epsscale{1.2} 
\plotone{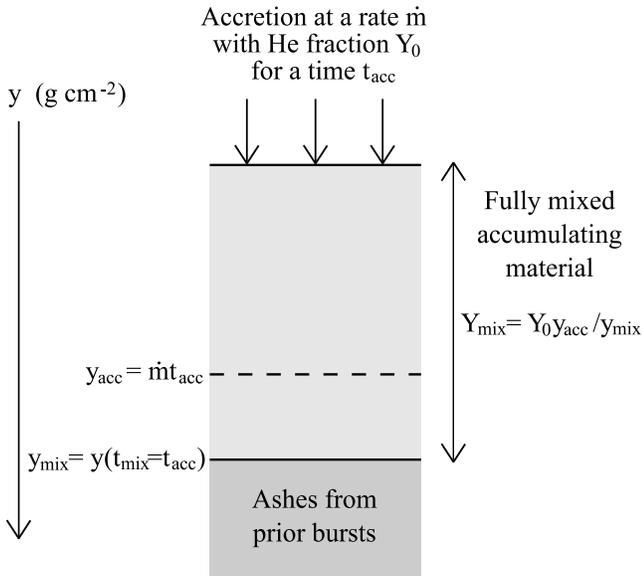}
\caption{Diagram demonstrating the main features of turbulent mixing.
Material mixes down to where $\tmix=\tacc$, which defines the depth
$\ymix$. The total amount of material that accretes is $\yacc=\dot{m}\tacc$,
giving a total accreted column of helium of $Y_0\yacc$. Within the mixed
layer the helium mass fraction is diluted to new mass fraction of
$Y_{\rm mix}=Y_0\yacc/\ymix$.}
\label{fig:diagram1}
\epsscale{1.0}
\end{figure}

   Once $\tmix<\tacc$ at the buoyancy barrier, the compositional profile of the
NS is very different, which we now explore. We treat the mixing as
complete, which we diagram in Figure \ref{fig:diagram1} and summarize
here. Material accretes at a rate $\dot{m}$ for a time $\tacc$ with a helium
mass fraction $Y_0$, supplying a column of material $\yacc=\dot{m}\tacc$. The
total column of helium that accretes during this time is therefore $Y_0\yacc$.
Mixing causes this newly accreted material to mix past $\yacc$ to the mixing depth
$\ymix$ defined as where $\tmix=\tacc$. The helium is fully mixed
down to this depth, resulting in a diluted mass fraction $Y_{\rm mix}=Y_0\yacc/\ymix$
within the mixed layer. In the following sections we consider
two scenarios that can result from the mixing: (1) unstable ignition of the mixed
fuel, and (2) stable burning when the material mixes to sufficient depths.


\subsection{Numerical Calculations of Mixed Ignition}
\label{sec:mixedignition}

   As the column of accreted material grows, it can still reach the correct conditions
for unstable ignition, but mixing causes two changes: (1) the ignited layer has
a diluted helium fraction, $Y_{\rm mix}<Y_0$, and (2) the ignition takes
place at the base of the mixed layer, at a depth $\ymix>\yacc$, resulting in a
recurrence time for ignition much less than when mixing is not included.

   Both of these effects are easiest to explore using a semi-analytic model. In this section
all calculations use $Y_0=1$. We consider other values of $Y_0$ for our analytic
estimates in the next section. We solve for the mixed accumulating structure
by first assuming an amount of accretion $\yacc$, which for a given $\dot{m}$
implies an accretion time $\yacc/\dot{m}$. We integrate the radiative diffusion
equation (eq. [\ref{eq:flux}]) down to a depth where $\tmix=\tacc$ giving $\ymix$,
where we assume a constant
flux profile as was discussed in \S \ref{sec:nomix} since little helium burning takes place
during accumulation. We then estimate the mixed
helium fraction as $Y_{\rm mix}=\yacc/\ymix$. This estimate is improved by substituting
$Y_{\rm mix}$ back into our envelope integration, and iterating until $Y_{\rm mix}$
converges. The non-helium component in the layer is taken to be iron.

\begin{figure}
\epsscale{1.2} 
\plotone{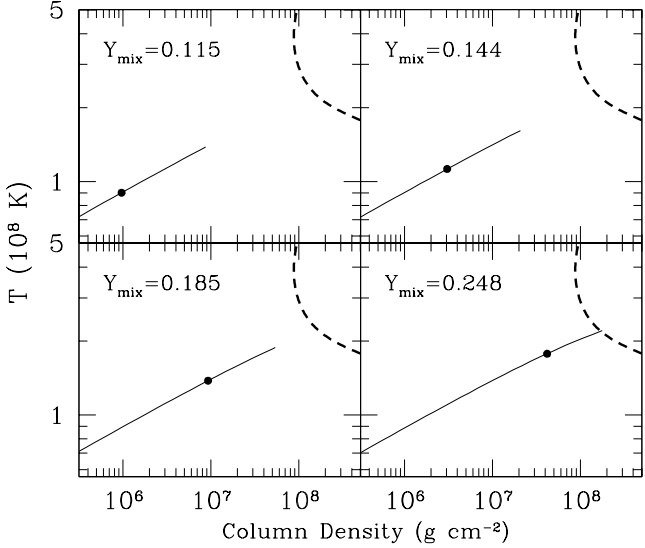}
\caption{The four panels show how the fully mixed accumulating layer evolves
in time until it reaches conditions necessary for unstable ignition. The parameters of
the NS are $\dot{m}=0.1\dot{m}_{\rm Edd}$ and $\Omega=0.1\Omega_{\rm K}$. In each
panel, the column of helium that has been accreted is denoted by a filled circle, which is from
left to right, and up to down $\yacc=10^6$, $3\times10^6$, $10^7$, and $4\times10^7\ {\rm g\ cm^{-2}}$.
 Mixing takes place down to the column reached by the thin solid line. The mixed helium fraction,
$Y_{\rm mix}$ is displayed in the upper left-hand corner of each panel. The ignition
curve associated with each $Y_{\rm mix}$ is shown as a thick dashed line.}
\label{fig:mixed_ignition}
\epsscale{1.0}
\end{figure}

   In Figure \ref{fig:mixed_ignition} we plot the resulting profiles for a NS accreting
at $\dot{m}=0.1\dot{m}_{\rm Edd}$ and $\Omega=0.1\Omega_{\rm K}$. In the four panels we
consider values of $\yacc$ of $10^6$, $3\times10^6$, $10^7$, and $4\times10^7\ {\rm g\ cm^{-2}}$
(from left to right and then up to down, denoted by the filled circles), which is meant to
mimic the accumulation of fuel
on the NS surface. For each integration the envelope profile continues down to a depth
$\ymix$, which also gives $Y_{\rm mix}$ as displayed in each panel. The last model
reaches the conditions necessary for ignition at the base of the mixed layer.
The recurrence time for these mixed-ignition models is much less than
for those without mixing. For the plotted model, the recurrence time is
$t_{\rm rec}=4\times10^7\ {\rm g\ cm^{-2}}/1.5\times10^3\ {\rm g\ cm^{-2}\ s^{-1}}\approx44\ {\rm minutes}$.
In contrast the model shown in Figure \ref{fig:accumulating} with the same accretion
rate of $\dot{m}=0.1\dot{m}_{\rm Edd}$
has $t_{\rm rec}\approx1.5\ {\rm days}$! The shorter recurrence time is not only due
to mixing carrying helium down to deeper depth, but
also a change in the thermal profile. Since significant iron is mixed up into the accumulating
material, the free-free opacity, which scales $\kappa_{\rm ff}\propto Z^2/A$, where $Z$ and $A$ are
the charge and mass per nucleon, respectively, is now the dominate opacity mechanism.
The accumulating layer is more opaque and therefore hotter for a given
flux in comparison to the pure-helium models considered before, which contributes
to the shallow ignition depths.
 
\begin{figure}
\epsscale{1.2} 
\plotone{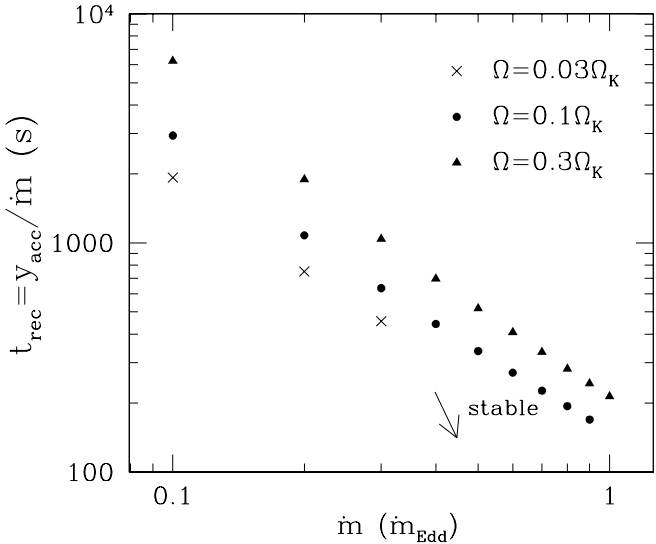}
\caption{The recurrence time for mixed-ignition models as a function
of $\dot{m}$. The symbols denoted different spins, as shown in the key.
Models that are at sufficiently high $\dot{m}$ or low $\Omega$ do not ignite
unstably, and thus are not plotted.}
\label{fig:trec}
\epsscale{1.0}
\end{figure}

   We calculate $t_{\rm rec}$ for a grid of models with various $\dot{m}$ and $\Omega$ in
Figure \ref{fig:trec}. The recurrence time is shorter for stronger mixing, which
occurs at high $\dot{m}$ or low $\Omega$, and can be as
short as $\approx5-30\ {\rm minutes}$. We warn
though that all of these calculations assume that complete mixing can occur, and
as we already showed in \S \ref{sec:nomix}, buoyancy may prevent this (eq. [\ref{eq:mdotcrit1}]).
Nevertheless, it is interesting to calculate the mixed-ignition conditions for a wide range of parameter
space because the conditions left from previous bursts may vary, and quantities such as
$\Delta\ln\mu$ may be smaller at times if, for example, there is incomplete burning
in a previous burst. At sufficiently high $\dot{m}$ or low $\Omega$ the ignition takes
place at high enough temperatures that the envelope does not ignite unstably.
We consider this case in more depth in the following section.

   The short recurrence times that we find are similar to some seen for 
multiple bursts \citep[][and references therein]{gal06}. However, our model cannot
explain the energetics of these bursts. The energetics are typically quantified
in a distance independent measure,  the so-called $\alpha$-value, which is the ratio of energy
released in the
persistent emission between bursts to the energy of the burst itself. For pure helium
ignition $\alpha\sim100$. In contrast, the $\alpha$-values for
the short recurrence time bursts are typically $\sim10$ \citep{boi07}.
Since the only helium that will burn in our mixed ignition
models is that accreted since the last outburst, we always find $\alpha\sim100$.
Therefore we still require an additional nuclear energy source, such as incomplete
burning from the previous X-ray burst, to explain such a low $\alpha$-value.
This in fact may not be a problem because
{\it incomplete burning may naturally explain the short recurrence times, since this would
lead to smaller compositional gradients and therefore stronger mixing.}

\begin{figure}
\epsscale{1.2} 
\plotone{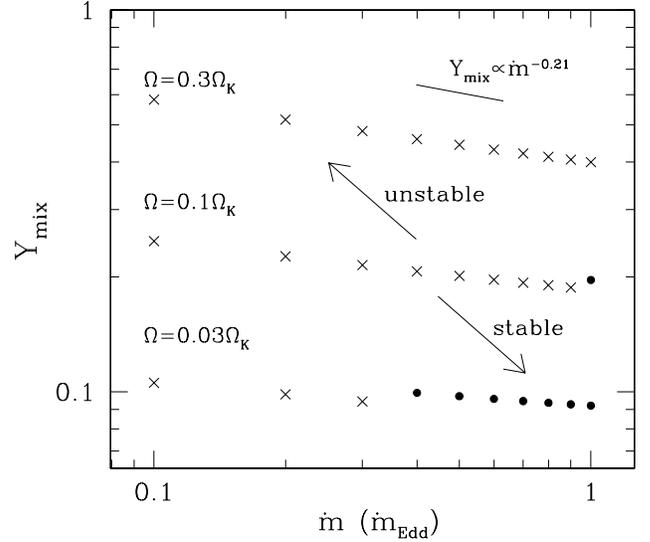}
\caption{The mixed helium mass fraction, $Y_{\rm mix}$ as a function of $\dot{m}$
and $\Omega$. The crosses are mixed ignition models. The filled circles are for
stably burning models and give the amount of helium present in the steady-state
mixing and burning layer. Steady-state burning therefore require
either high $\dot{m}$ or low $\Omega$. The scaling of $Y_{\rm mix}\propto\dot{m}^{-0.21}$
is derived in equation (\ref{eq:ymix2}) and is consistent with these numerical
results. Without mixing {\it all of the these considered models
ignite unstably} since stable accretion requires $\dot{m}\gtrsim10\dot{m}_{\rm Edd}$.}
\label{fig:summary}
\epsscale{1.0}
\end{figure}

   We plot the $Y_{\rm mix}$
as a function of $\dot{m}$ and $\Omega$ in Figure \ref{fig:summary}. The crosses correspond to
models that ignite unstably. The filled circles are stable accreting models that are discussed
on \S \ref{sec:steadystate} and \S \ref{sec:criticalmdot}. The implications of this mixed ignition
for the unstable burning during the X-ray burst can be easily tested by more sophisticated numerical
simulations {\it by just considering a mixed accumulating column}, for example, by
artificially setting $Y\approx0.1-0.6$ as shown in this Figure \ref{fig:summary}. Since
viscous heating is negligible, these initial tests do not need to resolve the shearing profiles.


\subsection{Analytic Estimates of Mixed Ignition}

   We now estimate the properties of the mixed ignition models.
These solutions directly show how mixing depends
on the properties of the accumulating layer, in particular the prefactor $\alpha_{\rm TS}$ and
$Y_0$, without having to consider a multitude of models. The effects of a free-free opacity
are important in deriving the correct atmospheric conditions. To include this analytically, we use the
free-free opacity from \citet{sch99}, simplified to a one-component plasma and with
the dimensionless Gaunt factor set to unity,
\be
	\kappa_{\rm ff}\approx3.77\ {\rm cm^2\ g^{-1}}\frac{\dens}{\temp^{7/2}}
		\frac{Z^2}{A},
\ee
where we have estimated $\mu_e\approx2$ as is correct within 10\% for the any of the elements of
interest. Integrating the radiative diffusion equation (eq. [\ref{eq:flux}]), assuming a constant flux of
$F=10^{21}\ {\rm ergs\ cm^{-2}\ s^{-1}}F_{21}$ and an ideal gas equation of state, the temperature
as a function of column $y$ is
\be
	T(y)=1.8\times10^8\ {\rm K}\ \lp\mu_{1.33}F_{21}Z^2/A\rp^{2/17}y_8^{4/17},
	\label{eq:temp_profile}
\ee
where $y_8\equiv y/10^8\ {\rm g\ cm^{-2}}$. Using equation (\ref{eq:tmix}), we
find the mixing timescale as a function of $y$,
\be
	\tmix = 3.5\times10^3\ {\rm s}\ \alpha_{\rm TS}^{-1}\mu_{1.33}^{-0.44}\lp F_{21}
		Z^2/A\rp^{-0.19}\mdot^{-1}\spin^{0.75}y_8^{1.37}.
		\nonumber
		\\
\ee
Note that $\tmix\propto y^{1.37}$ is a higher power than $\tacc=y/\dot{m}\propto y$.
This explicitly shows that mixing dominates at lower $y$, but accretion always wins at
some depth. Setting $\tmix$ equal to $\tacc=\yacc/\dot{m}$ gives the depth where mixing can extend
to for a given column of accreted material $\yacc$,
\be
	\ymix(\yacc)&=&1.6\times10^8\ {\rm g\ cm^{-2}}\alpha_{\rm TS}^{0.73}\mu_{1.33}^{0.32}
		\nonumber
		\\
		&&\times\lp F_{21}Z^2/A\rp^{0.14}\spin^{-0.55}y_{\rm acc,8}^{0.73},
		\label{eq:ymix1}
\ee
where $y_{\rm acc,8}=y_{\rm acc}/10^8\ {\rm g\ cm^{-2}}$. The mixed helium fraction down
to this depth is
\be
	Y_{\rm mix}(\yacc) &=&\frac{Y_0\yacc}{\ymix(\yacc)}
		\nonumber
		\\
		&=&0.63\ \alpha_{\rm TS}^{-0.73}Y_0\mu_{1.33}^{-0.32}
			(F_{21}Z^2/A)^{-0.14}\spin^{0.55}y_{\rm acc,8}^{0.27}.
			\nonumber
			\\
		\label{eq:he_frac1}
\ee
Since $Y_{\rm mix}\propto\yacc^{0.27}$ the strength of mixing decreases ($Y_{\rm mix}$
gets larger) as more material accreted, which was demonstrated by the four panels in Figure \ref{fig:mixed_ignition}.

   We next estimate what ignition depth is expected for this fully mixed accumulating layer.
The energy generation rate for triple-$\alpha$ burning is approximated as
\be
	\epsilon_{3\alpha}=5.3\times10^{23}\ {\rm ergs\ g^{-1}\ s^{-1}}\ f
		\frac{\dens^2Y_{\rm mix}^3}{\temp^3}\exp\lp-\frac{44}{\temp}\rp,
\ee
where $f$ is factor that accounts for screening effects. To make progress analytically
we expand the exponential as
$\exp\lp-44/\temp\rp\approx7.95\times10^{-10}(\temp/2.1)^{21}$. Using our temperature
profile (eq. [\ref{eq:temp_profile}]), the condition that $d\epsilon_{3\alpha}/dT=d\epsilon_{\rm cool}/dT$
(eq. [\ref{eq:stability}]) implies an ignition depth of
\be
	y_{\rm ign} &=& 9.4\times10^7\ {\rm g\ cm^{-2}}\ f^{-0.15}\mu_{1.33}^{-0.57}
		F_{21}^{-0.13}
		\nonumber
		\\
		&&\times(Z^2/A)^{-0.28}Y_{\rm mix}^{-0.44}.
\ee
Setting $y_{\rm ign}=\ymix$ from equation (\ref{eq:ymix1}), we solve for $\yacc$, the
critical column of material that must be accreted to cause ignition. This is then substituted
back into equations (\ref{eq:ymix1}) and (\ref{eq:he_frac1}) to find that ignition occurs
at a depth
\be
	y_{\rm mix,ign} &=& 1.2\times10^8\ {\rm g\ cm^{-2}}\ \alpha_{\rm TS}^{0.38}
		f^{-0.13}\mu_{1.33}^{-0.33}Y_0^{-0.38}
		F_{21}^{-0.039}
		\nonumber
		\\
		&&\times(Z^2/A)^{-0.17}\spin^{-0.29}.
		\label{eq:ymixign}
\ee
with a composition of
\be
	Y_{\rm mix,ign} &=& 0.57\ \alpha_{\rm TS}^{-0.86}f^{-0.047}\mu_{1.33}^{-0.56}Y_0^{0.86}
		F_{21}^{-0.21}
		\nonumber
		\\
		&&\times(Z^2/A)^{-0.25}\spin^{0.65}.
\ee
Comparing this with the numerical calculation is easiest if we assume $F$ is set by $\dot{m}$
($F_{21}=2.2\dot{m}_{0.1}$) as
well as scaling $Z^2/A\approx12$ and $\mu\approx2.1$ as appropriate for the iron-rich
composition. This gives
\be
	Y_{\rm mix,ign}\approx0.20\ \alpha_{\rm TS}^{-0.86}\mdot^{-0.21}\spin^{0.65}.
	\label{eq:ymix2}
\ee
The recurrence time is $t_{\rm rec}=Y_{\rm mix,ign}y_{\rm mix,ign}/\dot{m}$, resulting in
\be
	t_{\rm rec}\approx950\ {\rm s}\ \alpha^{-0.48}\mdot^{-1.25}\spin^{0.36}.
	\label{eq:trec}
\ee
Equations (\ref{eq:ymix2}) and (\ref{eq:trec})
confirm the scalings found for the numerical calculations in Figures \ref{fig:trec}
and \ref{fig:summary}. We have plotted the $Y_{\rm mix}\propto\dot{m}^{-0.21}$ scaling
in Figure \ref{fig:summary} to emphasize this. These analytic results also show
how strongly these results depend on the parameter $\alpha_{\rm TS}$, for which
$Y_{\rm mix,ign}$ is especially sensitive.


\subsection{Steady-State Mixing and Burning}
\label{sec:steadystate}

\begin{figure}
\epsscale{1.2} 
\plotone{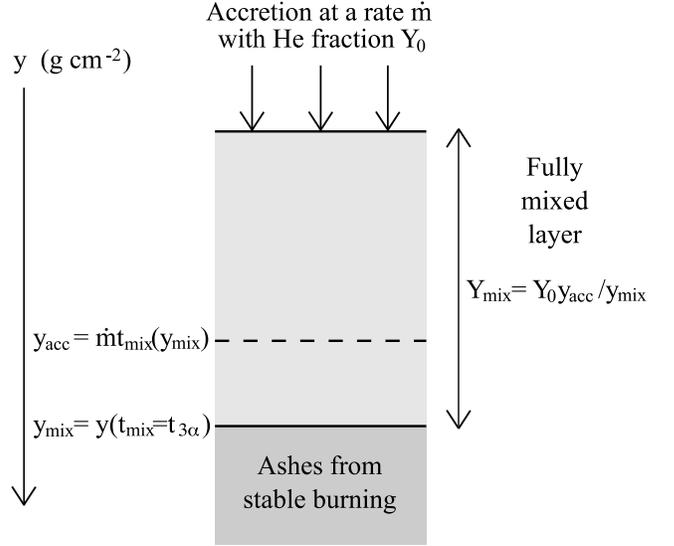}
\caption{Diagram demonstrating the main features of steady-state mixing and
burning. Material can mix further down to where $\tmix=\tnuc$, which defines
the depth $\ymix$. During a timescale $\tmix(\ymix)$ the amount of material
which has accreted is $\yacc=\dot{m}\tmix(\ymix)$. Within the mixed layer the
helium mass fraction is diluted to new mass fraction of $Y_{\rm mix}=Y_0\yacc/\ymix$.}
\label{fig:diagram2}
\epsscale{1.0}
\end{figure}

   Another possibility is that the helium is mixed and
burned by triple-$\alpha$ reactions in steady-state, leading to stable burning.
The basic idea is similar to that described above for mixed ignition, except now
the depth of the accumulating layer is set by the helium burning
timescale, $t_{3\alpha}=E_{3\alpha}/(Y\epsilon_{3\alpha})$, where
$E_{3\alpha}=5.84\times10^{17}\ {\rm ergs\ g^{-1}}$
is the energy per mass released from this burning.
As shown in Figure \ref{fig:diagram2}, material is mixed to
sufficient depths where $\tnuc$ (which decreases with depth)
is equal to $\tmix$ (which increases with depth), which defines a mixing (or burning) depth
$\ymix=y(\tmix=\tnuc)$. During a mixing timescale, the amount of material that is accreted
is $\yacc=\dot{m}\tmix(\ymix)$, so that the total column of helium that has been accreted
is $Y_0\dot{m}\tmix(\ymix)$. This is diluted over a depth $\ymix$, so that the mixed
helium fraction is $Y_{\rm mix}=Y_0\yacc/\ymix$. This is all occurring in steady-state,
material moves through the mixed layer at a rate $\dot{m}$, but
this layer does not move up or down in pressure (column) coordinates, as
the burning is stable.

   The basic results of steady-state mixing and burning are best shown using a
simple numerical model.
The equation that describes helium continuity, including depletion by triple-$\alpha$
burning, becomes, in the plane-parallel limit \citep{fuj93},
\be
	\frac{dY}{dt}=\frac{1}{R^2\rho}\frac{\partial}{\partial z}\lp R^2\rho D\frac{\partial Y}{\partial z}\rp
		+\lp \frac{\partial Y}{\partial t}\rp_{3\alpha}.
\ee
We assume changes in $\rho$ and $D$ with depth are small in comparison to changes in
$Y$, and, following our derivation of the angular momentum equation in \S \ref{sec:baroclinic},
we take the steady state limit to derive
\be
	-\frac{\dot{m}}{\rho}\frac{dY}{dz}=D\frac{d^2Y}{dz^2}-\frac{Y}{\tnuc}.
\ee
In the limit where $\tacc\gg\tmix$ that we are interested in, the term on the left-hand side
can be dropped. Finally, making the approximation that $\rho Dd/dz~\approx~y/\tmix$, we find
\be
	\frac{dY}{dy}=-\frac{\tmix}{\tnuc}\frac{Y}{y}.
	\label{eq:helium}
\ee
This equation mimics the properties we expect from mixing. When
mixing is strong $\tmix/t_{3\alpha}\ll1$, and $dY/dy\approx0$, the composition
does not change with depth. At the depth where
$\tmix\approx t_{3\alpha}$, $dY/dy\approx -Y/y$ and the helium is depleted exponentially.
All the helium burns into carbon, so that carbon has a mass fraction $X_{12}=1-Y$.

   The envelope profiles are found from simultaneously integrating three
differential equations: (1) radiative transfer, equation (\ref{eq:flux}),
(2) the entropy equation, $dF/dy=-\epsilon_{3\alpha}$, and
(3) continuity of helium, equation (\ref{eq:helium}). We integrate using a shooting
method, but first we must set three boundary conditions for the flux, temperature,
and helium mass fraction. Since all of the accreted helium must burn if the envelope
is in steady-state, we set the surface flux to
$F=Y_0E_{3\alpha}\dot{m}+F_c$, where
$F_c=150\ {\rm keV\ nuc^{-1}}\langle\dot{m}\rangle$
(as discussed in \S \ref{sec:shearprofile}).
The surface temperature is set from the radiative zero solution \citep{sch58}.
The helium abundance in the mixed region, $Y_{\rm mix}$, is an eigenvalue. It is
varied until shooting gives the correct base flux of $F_c$.
This is easily found through iteration since when $Y_{\rm mix}$
is set too large, too much burning occurs, and the base flux is too small
(and vice versa for small $Y_{\rm mix}$).

\begin{figure}
\epsscale{1.3} 
\plotone{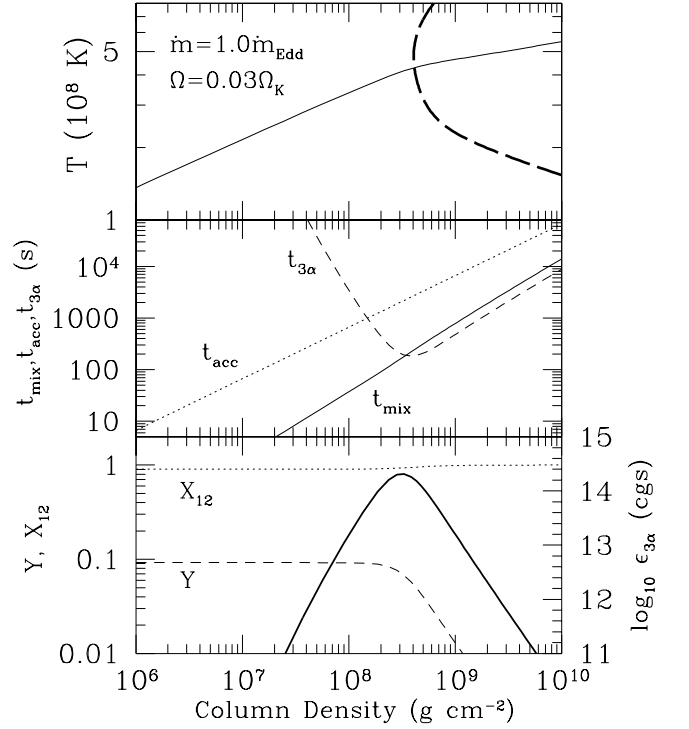}
\caption{An example steady-state mixing and burning envelope model using
$Y_0=1$, $\dot{m}=1.0\dot{m}_{\rm Edd}$, $\Omega=0.03\Omega_{\rm K}$, and
$\alpha_{\rm TS}=1$. The material within the mixed layer has $Y_{\rm mix}=0.097$.
The top panel shows the temperature profile ({\it solid line}), as well
as the critical curve ignition using a helium mass fraction of $0.052$
({\it thick dashed line}, which
is the helium mass fraction at the burning depth).
The middle panel compares the key timescales.
The bottom panel shows the helium ({\it dashed line})
and carbon abundances ({\it dotted line}), as well
as the energy generation rate for helium burning
({\it thick solid line}).}
\label{fig:steady}
\epsscale{1.0}
\end{figure}

   In Figure \ref{fig:steady} we plot a steady-state
envelope using $Y_0=1$, $\dot{m}=0.3\dot{m}_{\rm Edd}$,
$\Omega=0.1\Omega_{\rm K}$, and $\alpha_{\rm TS}=1$.
The shooting method demonstrates that the initial helium abundance within
the mixed layer is $Y_{\rm mix}=0.097$. The top panel shows the
temperature profile ({\it thin solid line}), and the critical curve
for stability where $d\epsilon_{3\alpha}/dT=d\epsilon_{\rm cool}/dT$
({\it thick dashed line}). The bottom panel shows that the
majority of triple-$\alpha$ burning occurs at
$\approx3\times10^8\ {\rm g\ cm^{-2}}$ ({\it thick solid line}), which is where
the helium is depleted. Comparing
the middle and bottom panel shows that the majority of the burning takes place
near where $\tmix=t_{3\alpha}$ (as required by construction),
which is deeper than where $\tacc=t_{3\alpha}$
(the normal condition for steady burning).

\begin{figure}
\epsscale{1.3} 
\plotone{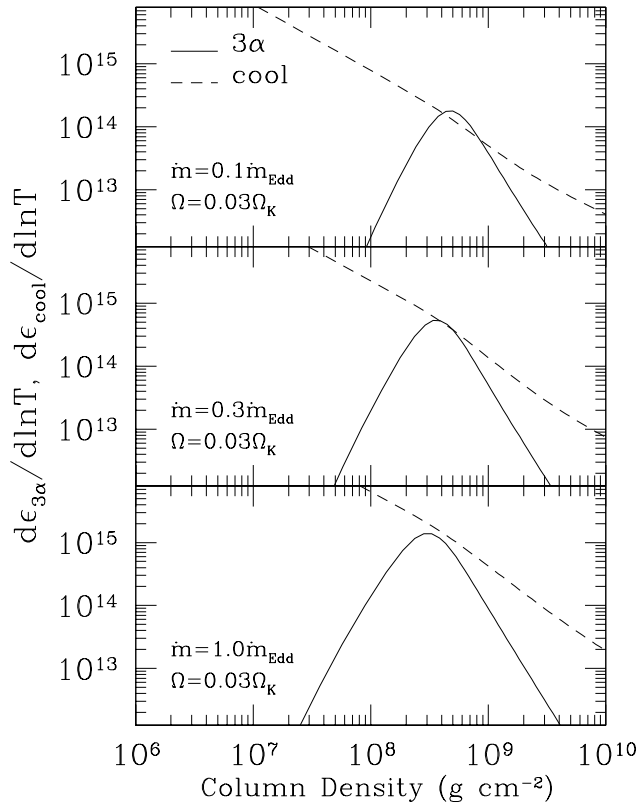}
\caption{A comparison of $d\epsilon_{3\alpha}/d\ln T$
({\it solid lines})
and $d\epsilon_{\rm cool}/d\ln T$ ({\it dashed lines}) as a function of depth for
$\dot{m}=0.1$, $0.3$, and $1.0\dot{m}_{\rm Edd}$ ({\it top to bottom
panel}). All the models use $\Omega=0.03\Omega_{\rm K}$,
$\alpha_{\rm TS}=1$, and $Y_0=1$. This demonstrates that only the
$\dot{m}=1.0\dot{m}_{\rm Edd}$ model is stable out of these three.}
\label{fig:stability}
\epsscale{1.0}
\end{figure}
   To make sure the steady-state models we find are physically realizable, we must check
the thermal stability of the helium burning.
In Figure \ref{fig:stability} we compare the quantities $d\epsilon_{3\alpha}/d\ln T$
and $d\epsilon_{\rm cool}/d\ln T$ for three different accretion rates. If
the cooling derivative is always larger, then the model is thermally stable.
We find stable accretion at $\dot{m}$'s considerably less than the stable
accretion rate of $\dot{m}\approx10\dot{m}_{\rm Edd}\approx2\times10^6\ {\rm g\ cm^{-2}\ s^{-1}}$
expected for pure helium
accretion estimated without mixing \citep{bil95,bil98a}.
Models that are found to be stable in this way are plotted as filled circles
in Figure \ref{fig:summary} (from \S \ref{sec:mixedignition}).
Only at large $\dot{m}$
and small $\Omega$ are the models found to be stable. If we where to increase $\alpha_{\rm TS}$,
a wider range of the models in Figure \ref{fig:summary} become stable.

\subsection{The Critical $\dot{m}$ for Stability}
\label{sec:criticalmdot}

   Since we have found a set of models that can stably accrete, mix, and then burn
helium, it is interesting to ask what accretion rates and spins are required for this to occur, and how
does it depend on parameters such as $\alpha_{\rm TS}$.

   First we must derive the correct condition for stability including the fact that
free-free opacity is important in setting the radiative profile. The one-zone condition for
stability at the base of the mixed layer is
\be
	\frac{d\epsilon_{3\alpha}}{d\ln T}<\frac{d\epsilon_{\rm cool}}{d\ln T},
\ee
where the derivatives are taken at constant pressure. For $\epsilon_{3\alpha}\propto T^\zeta\rho^\chi$,
where $\zeta=44/\temp-3$,
stability requires \citep{bil98a}
\be
	\zeta-4+\frac{\partial\ln\kappa}{\partial\ln T}+\frac{\partial\ln\rho}{\partial\ln T}
	\lp \chi+\frac{\partial\ln\kappa}{\partial\ln\rho}\rp<0.
\ee
Substituting the scalings for a free-free opacity and ideal gas equation of state
we find $\temp>3.26$ is required for stability.

   By substituting the analytic form we found for the mixed ignition depth (eq. [\ref{eq:ymixign}])
into the temperature profile (eq. [\ref{eq:temp_profile}]) we can find the temperature at
the base of the mixed layer,
\be
	T_{\rm ign}=2.54\times10^8\ {\rm K}\ \alpha_{\rm TS}^{0.09}\mdot^{0.11}\spin^{-0.067}.
\ee
By simply asking when $T_{\rm ign}>3.26\times10^8\ {\rm K}$, we derive
a stabilizing $\dot{m}$ of
\be
	\dot{m}_{\rm crit,2}=1.0\dot{m}_{\rm Edd}\alpha_{\rm TS}^{-0.83}\spin^{0.62}.
	\label{eq:mdotcrit2}
\ee
This is in reasonable agreement with Figure \ref{fig:summary}, which shows stability
occurs for $0.9\dot{m}_{\rm Edd}\lesssim\dot{m}\lesssim1.0\dot{m}_{\rm Edd}$ for
$\Omega=0.1\Omega_{\rm K}$. It is interesting that $\dot{m}_{\rm crit,2}$
is near (within an order of magnitude) a value where bursts are observed to change. It is conceivable
that this mechanism may act to stabilize X-ray bursts for the hydrogen-rich accreting systems,
and given the strong scaling $\dot{m}_{\rm crit,2}$ has with $\alpha_{\rm TS}$ it is possible
that more detailed calculations could give results that agree even better with
the critical accretion rates that are observed.
We discuss this idea in more detail in the following section.


\section{Discussion and Conclusion}
\label{sec:theend}

\begin{figure}
\epsscale{1.2} 
\plotone{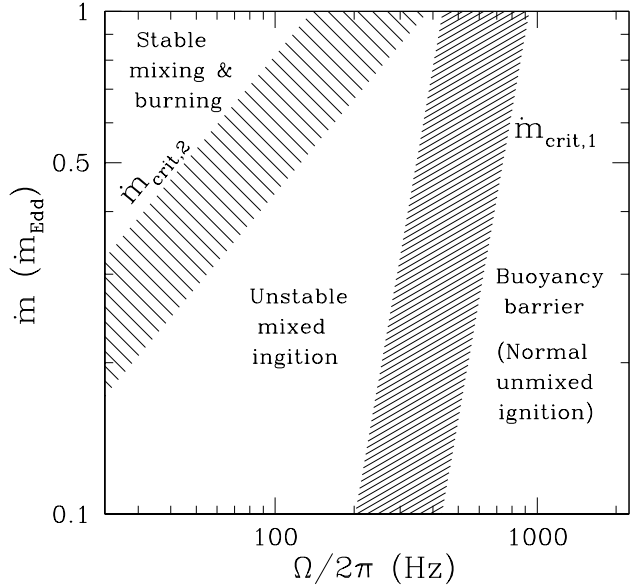}
\caption{Summary of the three regimes of burning found for models including
mixing. The boundaries between each regime are shown as shaded regions to
emphasize possible uncertainty in the strength of mixing (we consider
$\alpha_{\rm TS}=0.7-1.5$).
The heavy shaded region divides where the buoyancy barrier is overcome
($\dot{m}_{\rm crit,1}$, eq. [\ref{eq:mdotcrit1}]), and the light shaded region divides
between stable and unstable mixed burning ($\dot{m}_{\rm crit,2}$, eq. [\ref{eq:mdotcrit2}]).
An additional uncertainty in $\dot{m}_{\rm crit,1}$ is the value of $\Delta\ln\mu$, which
could vary depending on the results of previous bursts.}
\label{fig:phase}
\epsscale{1.0}
\end{figure}

   We have revisited the problem of angular momentum transport in the surface
layers of accreting NSs. We found that the hydrodynamic instabilities used
by \citet{fuj93} in a previous study are dwarfed by the magnetic effects of
the Tayler-Spruit dynamo. The large viscosity provided by this process results
in a very small shear rate and negligible viscous heating.
The turbulent mixing is sufficiently large to have important consequences
for X-ray bursts. We constructed simple models, both analytic and numerical,
to explore mixing for pure helium accretion. From these models we can make
a few conclusions that are likely general enough to apply to most viscous mechanisms.
As a guide, we show the different burning regimes
we find in Figure \ref{fig:phase}. These can be summarized as follows:
\begin{itemize}
\item Mixing is strongest at large $\dot{m}$ (when angular momentum is
being added at greater rates) and small $\Omega$ (which gives a larger
relative angular momentum between the NS and accreted material).
\item Mixing has trouble overcoming the buoyancy barrier at chemical
discontinuities. But once mixing breaks through this, it extends down
to a depth where $\tmix=\tacc$, which is generally much deeper than the
accreted column. This means that mixing should turn
on abruptly (as a function of either $\dot{m}$ or $\Omega$). It also means
that the importance of mixing depends on the particular ashes left over
from previous bursts. If, for example, incomplete burning results in small compositional
gradients, mixing would be important in subsequent bursts.
\item Mixing of freshly accreted material with the ashes from previous X-ray
bursts can lead to two new effects. First, the layer may
ignite, but now in a mixed environment with a short recurrence time
of $\sim5-30\ {\rm minutes}$. Second, if the mixing
is strong enough, accreted helium can mix and burn in steady-state,
quenching X-ray bursts. Both of these regimes have observed analogs, namely the
short recurrence time bursts \citep[for example in,][]{boi07} and the stabilization of bursting seen at
$\approx0.1\dot{m}_{\rm Edd}$ \citep{cor03}.
\end{itemize}
The mixed ignition case can be studied easily using the current
numerical experiments \citep[e.g.,][]{woo04} by just artificially accreting
fuel with a mixed composition of $Y_{\rm mix}\approx0.1-0.6$. These calculations are simplified by
our conclusion that shearing and heating can be ignored, at least for initial
studies. We next conclude by speculating about some
of the other ramifications of turbulent mixing.


\subsection{Superbursts}

   An ongoing mystery in the study of bursting NSs is the recurrence times
for superbursts, thermonuclear ignition of carbon in the X-ray burst ashes at
columns of $\approx10^{11}-10^{12}\ {\rm g\ cm^{-2}}$ \citep{cb01,sb02}.
This problem could be alleviated by enhanced heating from the core on the
order of $1\ {\rm MeV\ nuc^{-1}}$ \citep{cum06}, but this is more than is
expected theoretically, even in the newest calculations \citep{gup06}.
Shear heating is not large enough to solve this problem, as
demonstrated in Figures \ref{fig:transport1} and \ref{fig:transport2}.

   The regime of stable helium burning we have found may, however, create a carbon
rich ocean that would assist in the ignition of superbursts. Carbon fractions of
greater than 10\% are needed to reproduce the lightcurves and recurrence
times of superbursts \citep{kee06}. Calculations of rp-process burning show that
{\it unstable} burning cannot give carbon fractions this high \citep{sch03}.
Observationally, \citet{int03} showed that the $\alpha$-value (see \S \ref{sec:mixedignition})
is preferentially large for superbursting systems, indicating that some stable
burning is occurring, perhaps due to the turbulent mixing we have studied.


\subsection{Hydrogen-rich Accretion}

   One of the main deficiencies of our calculations are the simplified compositions,
since most accreting NSs are expected to be accreting a fuel abundant in hydrogen.
If the only crucial burning is triple-$\alpha$, such envelopes can be considered
within the framework of our models by using a solar value of $Y_0=0.3$. Our models fail, though,
when sufficient carbon is produced by triple-$\alpha$ to feedback into hydrogen burning
\citep[which burns via the hot-CNO cycle,][]{hf65}.
Such a scenario is interesting thought because it could potentially
produce very hydrogen poor bursts.

   In the standard theoretical framework for X-ray bursts, flashes should
be hydrogen poor at low $\dot{m}$ (when there is sufficient time
for the hot-CNO cycle to act), and then become mixed hydrogen-helium fuel at higher
$\dot{m}$. Paradoxically, observations show a transition at a seemingly
universal luminosity of $\approx2\times10^{37}\ {\rm ergs\ s^{-1}}$
(approximately $0.1\dot{m}_{\rm Edd}$), but in the opposite sense
\citep{cor03}. \citet{cn06b} argue that their models in fact show this transition
\citep[also see][]{nh03} because carbon created during helium simmering
increases the the hot-CNO rate and the temperature, which decreases the
temperature sensitivity of triple-$\alpha$ reactions. To explain the discrepancies between
their models and more detailed numerical simulations \citep{woo04,heg05} requires a decrease
in the breakout reactions rate
of $^{15}$O$(\alpha,\gamma)^{19}$Ne \citep{cn06a,fis06}. Unfortunately, the most recent
experimental results do not support a decreased rate
\citep{tan07,fis07}. Furthermore,
helium accreting systems show this same transition in bursting properties, which
is not explained within their framework.

   Another idea that may recreate this trend is that
the fraction of the star covered by the fuel increases with the {\it global}
accretion fast enough that the {\it local} accretion rate actually decreases
\citep{bil00}. This interpretation is supported by the recent work of
\citet{heg05}, who claim that the mHz oscillations observed at around this
same universal luminosity of $2\times10^{37}\ {\rm ergs\ s^{-1}}$
\citep{rev01} are a probe of the local accretion
rate where bursting is occurring. The main problem with this suggestion is
that it is difficult to understand how the covering of the surface can remain
so anisotropic all the way down to the depths of where ignition occurs.
\citet{is99} calculated the spreading of accreted material from the equator,
where the viscosity is due to a turbulent boundary layer, and find that
spreading occurs orders of magnitude shallower in depth than where ignition takes place.

   Since mixing gets stronger with $\dot{m}$, it may be that
the observed transition is in fact hydrogen being turbulently mixed and
burned analogous to what we have found for pure-helium accretion.
A possible implication of such an explanation is that slowly spinning NSs are more
likely to have their hydrogen depleted resulting in helium-rich bursts.
Assuming that the burst oscillation frequencies are indicative of the NS spin
frequency
\citep[which is close to true or exactly true for all current explanations][]{hey04,lee04,pb05,pm06},
the systems can be broken into slow spinning ($\sim300\ {\rm Hz}$) and fast
spinning ($\sim600\ {\rm Hz}$) classes. This slowly spinning class includes
4U $1916-053$ \citep[$270\ {\rm Hz}$,][]{gal01},
4U $1702-429$ \citep[$330\ {\rm Hz}$,][]{mar99}
and 4U $1728-34$ \citep[$363\ {\rm Hz}$,][]{sm99}.
Do these systems show helium-rich looking bursts, and if so, is this due to turbulent
mixing destroying the hydrogen they are accreting? This can be answered by
looking at the recent summary of {\it RXTE} burst observations by \citet{gal06}.
4U $1916-053$ has bursts consistent with helium-rich fuel, but this system also
has an orbital period of $\approx50\ {\rm min}$ \citep{gri88}. This is an
``ultracompact'' system in which the donor is too small to be a H-rich star, so
the accretion is probably helium-rich to begin with. The other two NSs both
have bursts with decay times and $\alpha$-values that suggest helium-rich
fuel, which is also supported by model fits to radius expansion bursts from 4U $1728-34$
\citep{gal06}. Furthermore, the bursts of 4U $1728-34$ have look very similar
to those of 4U $1820-30$ \citep{cum03}, a known ultracompact
\citep[see discussion in][and references therein]{pod02}.
While it is possible that 4U $1702-429$ and 4U $1728-34$
have helium-rich donors (and are thus ultracompacts),
it may also be that they are accreting hydrogen-rich
fuel and show helium-rich bursts because their low spins lead to mixing.
As binary parameters of these systems
become better known, it will be more clear whether turbulent mixing is
indeed needed to explain their burst properties.

\acknowledgements
We thank Henk Spruit for helpful discussions.
This work was supported by the National Science Foundation
under grants PHY 99-07949 and AST 02-05956.


\begin{appendix}

\section{Mixing and Energy Conservation for the Tayler-Spruit Dynamo}

   In \S \ref{sec:tssummary} we summarized the prescriptions that \citet{spr02} provides
for the Tayler-Spruit dynamo. The mixing diffusivity is assumed to be equal to the
turbulent magnetic diffusivity. Although this is plausible, it is not shown rigorously.
Below we argue that such a scaling is consistent with energy conservation.

   Consider a layer differentially rotating with a speed $\Delta V$.
The equation for the energy per unit mass expresses the fact that energy
from differential rotation can go into either viscous shearing or vertical
mixing
 \be
 	\frac{d}{dt}\left[\frac{1}{2}(\Delta V)^2\right]
	&=& -\frac{1}{2}\nu(q\Omega)^2- \frac{2\Delta\rho gH}{\rho t_{\rm mix}},
	\nonumber
	\\
	&=& -\frac{1}{2}\nu(q\Omega)^2- 2N^2 D,
\ee
where $D=H^2/t_{\rm mix}$ is the mixing
diffusivity. The ratio of the two right hand terms is called the flux Richardson number
(Fujimoto 1988)
\be
	Rf = \frac{4N^2D}{\nu(q\Omega)^2}=\frac{4D}{\nu}Ri,
\ee
where $Ri\equiv N^2/(q^2\Omega^2)$ is the Richardson number. The flux
Richardson number is interpreted as the ratio of energy that goes into
mixing versus heating (and usually taken to be $Rf\sim0.1-1$).
From this result we can solve for the mixing diffusivity
\be
	D=\frac{Rf}{4Ri}\nu.
\ee
This matches (up to a factor of order unity, $Rf/4$) what Spruit (2002) gives for the mixing
diffusivity. The factor of $Ri$ in the denominator takes into account the difficulty in overcoming
buoyancy to do mixing. The larger the buoyancy, the larger $Ri$ is and the smaller the mixing
diffusivity.

\end{appendix}


\end{document}